\documentclass[prd,showpacs,showkeys,amssymb]{revtex4}
\usepackage{graphicx}
\usepackage{dcolumn}
\usepackage{bm}
\usepackage{extarrows}
\usepackage{color}
\usepackage{multirow}

\begin{document}

\title{$K\pi$ interaction in finite volume and the $K^*$ resonance}

\author{Dan Zhou}
\author{Er-Liang Cui}
\author{Hua-Xing Chen}
\email{hxchen@buaa.edu.cn}
\author{Li-Sheng Geng}
\author{Li-Hua Zhu}
\affiliation{School of Physics and Nuclear Energy Engineering and International Research Center for Nuclei and Particles in the Cosmos, Beihang University, Beijing 100191, China}

\begin{abstract}
We evaluate energy levels of the $K \pi$ system in the $K^*$ channel in finite volume using chiral unitary theory. We use these energy levels to obtain $K \pi$ phase shifts, and then obtain the $K^*$ mass and its decay width. We investigate their
dependence on the pion mass and compare this with Lattice QCD calculations. We also compare our method with the standard L\"uscher approach, and solve the inverse problem to obtain the $K \pi$ phase shifts from these ``synthetic'' lattice data.
\end{abstract}


\date{\today}

\maketitle
\section{Introduction}

Lattice QCD is developing very fast in these years. One can use this method to evaluate the discrete energy levels of the finite box, and then reconstruct phase shifts of the decay products in the continuum. To do this, one usually uses the L\"uscher's approach~\cite{luscher,Luscher:1990ux}, which has a higher accuracy and consistency with the decay channels of the hadrons, and so it is widely used in lattice studies~\cite{Bernard:2008ax,Bernard:2010fp}. These discrete energy levels can not be directly measured in the experiments. However, in Ref.~\cite{Doring:2011vk} the authors proposed one method to estimate them through an effective approach whose parameters are obtained by fitting the experimental data.

This method has been applied in Ref.~\cite{Doring:2011ip} to obtain finite volume results from the J\"ulich model for meson baryon interaction, and in Ref.~\cite{MartinezTorres:2011pr} to study the interaction of the $DK$ and $\eta D_s$ systems where the $D_{s0}^*(2317)$ resonance is dynamically generated from the interaction of these channels~\cite{Kolomeitsev:2003ac,Hofmann:2003je,Guo:2006fu,Gamermann:2006nm}; the case of the $\kappa$ resonance in the $K \pi$  $S$-wave channel is studied in Ref.~\cite{Doring:2011nd}; the case of $\Lambda_c(2595)$ resonance in the $DN$ and $\pi \Sigma_c$ channels in finite volume is studied in Ref.~\cite{Xie:2012np}. An extension of the approach of Ref.~\cite{Doring:2011vk} to the case of interaction of unstable particles is studied in Ref.~\cite{Roca:2012rx}. We also use it to study the interaction of two pions in the $\rho$ channel in finite volume~\cite{Chen:2012rp}.

In the present work we shall study the $K \pi$ interaction in the $K^*$ channel in finite volume. This $K^*$ meson has been measured very well in the experiments and we can use the chiral unitary model to well describe it. Recently several Lattice groups also studied it and evaluated the relevant discrete energy levels using the L\"uscher's approach~\cite{Fu:2012tj,Lang:2012sv,Dudek:2014qha}. Again we note that these energy levels can not be directly measured in the experiments, so one needs to make extra efforts in order to compare these energy levels with the experimental data of the $K^*$ meson. Lattice theorists usually transform these energy levels into the phase shifts, and then calculate the physical quantities of the $K^*$ meson. Accordingly, one can do the opposite process~\cite{Doring:2011vk}, and this is what we shall study in this paper, i.e., in this paper we shall follow the approach of Ref.~\cite{Doring:2011vk}, and inversely transform the experimental data of $K^*$  meson into ``synthetic'' energy levels. To do this we need to use the chiral unitary model to study the $K \pi$ interaction in the $K^*$  channel in finite volume. To make a complete analysis, we shall also use these ``synthetic'' data to calculate the phase shifts and then calculate the physical quantities for the $K^*$  meson. We shall refer to the results of Refs.~\cite{Fu:2012tj,Lang:2012sv,Prelovsek:2013ela} for comparison along the paper. We shall also compare our method with the standard L\"uscher approach, and solve the inverse problem and obtain the $K \pi$ phase shifts from these ``synthetic'' lattice data.

This paper is organized as follows. In Sec.~II we study the $K \pi$ scattering in the $K^*$ region using the chiral unitary model both in infinite space and in finite volume. Then in Sec.~III we use these formulae to evaluate energy levels and phase shifts. The pion mass dependence of these results is studied in Sec.~IV where we also study their comparison with the Lattice data. We compare our method with the standard L\"uscher approach in Sec.~V, and solve the inverse problem to obtain $K \pi$ phase shifts from these ``synthetic'' lattice data in Sec.~VI. Finally we show some concluding remarks in Sec.~VII.

\section{The Chiral Unitary Approach In Infinite and Finite Box}

The $K \pi$ scattering amplitude in $P$-wave has been studied in Refs.~\cite{Oller:1998zr,Xiao:2013mn} by using the chiral unitary model. In this paper we shall follow the same approach and use the following Bethe-Salpeter equation in their on-shell factorized form~\cite{Oller:1998zr,Xiao:2013mn,Oller:1997ti,Oller:2001fj} (for a quantitative study of off-shell effects in this context, see, e.g., Ref.~\cite{Altenbuchinger:2013gaa}):
\begin{eqnarray}
T(s) &=& {V(s) \over 1 - V(s) G(s)} \, .
\label{bethesal}
\end{eqnarray}
Here we only consider the $K \pi$ channel, but the $K \eta$ and $K \eta^\prime$ channels may also be important. In Ref.~\cite{Guo:2011pa}, the $K^*$(892) is studied with the coupled channels $K \pi$, $K \eta$ and $K \eta'$. The coupling of the $K^*$(892) to $K \pi$ is the dominance, but some smaller, although not negligible couplings to $K \eta$ and $K \eta'$ are also found. The couplings by themselves do not give a measure of the relevance of the channel, because if the mass of the channel is far away from the pole, the relevance would be much smaller for a same coupling. Furthermore, in such a case, the effect of these channels and other missing channels can be absorbed in the study with one channel by changing the subtraction constants, and the energy dependence of the potential a bit, which is explicitly done in our model. Indeed, the fit to the data with just the $K \pi$ channel is very good, as found in Ref.~\cite{Oller:1998zr} and shown below. Moreover, the elimination of one channel in terms of an effective potential for another channel in the content of lattice QCD analysis has been shown to be a valid and useful tool in Ref.~\cite{sasa}.

\begin{figure}[hbt]
\begin{center}
\includegraphics[scale=0.8]{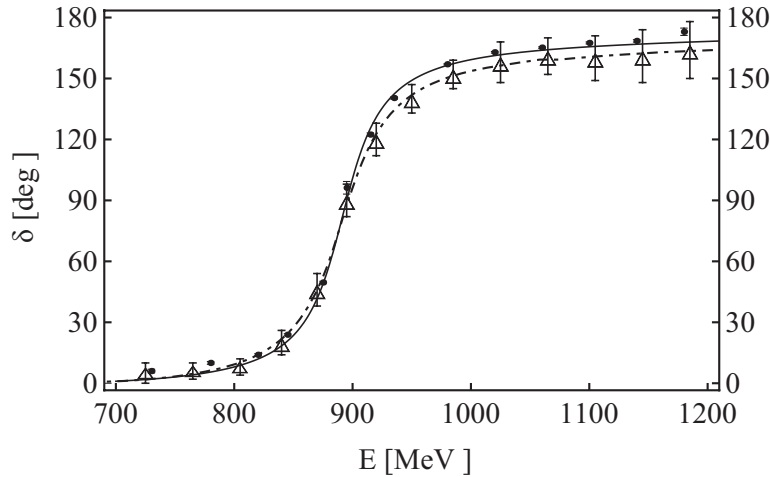}
\caption{The solid curve shows $K \pi$  scattering $P$-wave phase shifts obtained using Eq.~(\ref{bethesal}) and Eq.~(\ref{eq:delta}), and the dotdashed curve the results from Ref.~\cite{Xiao:2013mn}. The experimental data are taken from Ref.~\cite{Estabrooks:1977xe} and Ref.~\cite{Mercer:1971kn}, shown using  solid circles and triangles, respectively.}
\label{fig:fittingdata}
\end{center}
\end{figure}

The relevant $V$-matrix for the $K \pi$  scattering has been studied in Refs.~\cite{Oller:1998zr,Oller:2000ug,Xiao:2013mn}:
\begin{equation}
V(s) = -\frac{p^2}{2f^2}(1+\frac{2G_{V}^{2}}{f^2}\frac{s}{M_{K^*}^{2}-s}) \, ,
\label{eq:Vmatrix}
\end{equation}
where $M_{K^*}$ is the bare $K^*$ mass, $f$ is the $\pi/K$ decay constant, and $G_V$ is the coupling for a vector meson to two pseudoscalar mesons. We note that this potential $V(s)$ is a bit different from the one used in Ref.~\cite{Xiao:2013mn}, where the factor $p^2$ is absorbed into their $G$-function so that $V(s)$ does not depend on the momentum. The $G$-function for the two-meson ($\pi$-$K$) propagator having masses $m_\pi$ and $m_K$ is defined as
\begin{eqnarray}
\label{eq:gfunction}
G (p^2) &=& i \int {d^4 q \over (2 \pi)^4} {1 \over q^2 - m_\pi^2 + i \epsilon} {1 \over (p - q)^2 - m_K^2 + i \epsilon} \, ,
\end{eqnarray}
where $p$ is the four-momentum of the external meson-meson system. There are many methods to regularize this loop-function. In Ref.~\cite{Xiao:2013mn} the authors use the cut-off method, but in this paper we shall use the dimensional regularization which is more convenient when studying the $K \pi$  interaction in finite volume. We note that these two methods are equivalent up to certain energy level range, as proved in Ref.~\cite{Oller:2001fj}. The dimensional regularization result is
\begin{equation}
\begin{split}
G(s)=&\frac{1}{(4\pi)^2}\{a(\mu)+\log\frac{m_\pi^2}{\mu^2}+\frac{m_K^2-m_\pi^2+s}{2s}\log\frac{m_K^2}{m_\pi^2}\\
&+\frac{Q(\sqrt{s})}{\sqrt{s}}[\log(s-(m_K^2-m_\pi^2)+2\sqrt{s}Q(\sqrt{s}))+\log(s+(m_K^2-m_\pi^2)+2\sqrt{s}Q(\sqrt{s}))\\
&-\log(-s+(m_K^2-m_\pi^2)+2\sqrt{s}Q(\sqrt{s}))-\log(-s-(m_K^2-m_\pi^2)+2\sqrt{s}Q(\sqrt{s}))]\}\, ,
\end{split}
\label{eq:GDR}
\end{equation}
where $s=p^2$, $Q(\sqrt{s})$ is the on-shell momentum of the particles, $\mu$ is a regularization scale and $a(\mu)$ is a subtraction constant. In this paper we shall work in the center-of-mass frame, where the energy of the system is $E=\sqrt{s}$. The regularization parameters are chosen to be
\begin{eqnarray}
a(\mu) &=& - 1.0 \, ,
\\ \mu &=& M_{K^*} \, .
\end{eqnarray}
The two parameters $f$ and $G_V$ are taken from Ref.~\cite{Xiao:2013mn}:
\begin{eqnarray}
G_V &=& 53.81~{\rm MeV} \, ,
\\ f &=& 86.22~{\rm MeV} \, ,
\end{eqnarray}
but the parameter $M_{K^*}$ is a bit different from the one used in Ref.~\cite{Xiao:2013mn}, because we are using the dimensional regularization other than the cut-off method used in Ref.~\cite{Xiao:2013mn}. To fix $M_{K^*}$, we use the experimental data of the $K \pi$  $P$-wave phase shifts, which are related to the $T(s)$ through:
\begin{eqnarray}\label{eq:delta}
T(E) &=& { - 8 \pi E \over p \cot\delta(p) - i p } \, ,
\end{eqnarray}
where $p$ is the center-of-mass momentum.
We use the experimental data of Refs.~\cite{Mercer:1971kn,Estabrooks:1977xe}, and evaluate $M_{K^*}$. The fitting results are shown in Fig.~\ref{fig:fittingdata}, where $M_{K^*}$ is fitted to be:
\begin{eqnarray}
M_{K^*} &=& 919.03~{\rm MeV} \, .
\end{eqnarray}

\begin{figure}[hbt]
\begin{center}
\includegraphics[scale=0.8]{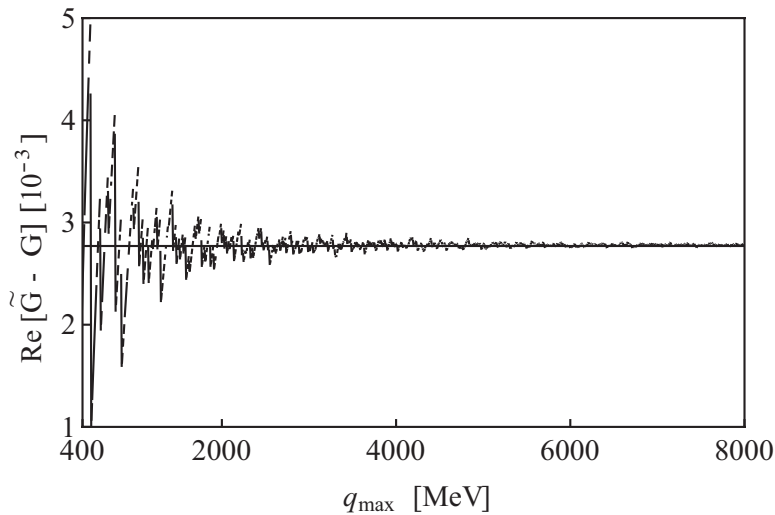}
\caption{The real part of Eq.~(\ref{gtilde}). Here we choose $L = 2.5~m_\pi^{-1}$ and $E = 800$ MeV.} \label{fig:difference}
\end{center}
\end{figure}

All the above formulae are defined in the infinite space. To study the $K^*$ meson in the finite volume, we simply change the $G$-function of dimensional regularization (Eq.~(\ref{eq:GDR})) by the one which is defined in the finite box of side $L$~\cite{Doring:2011jh,MartinezTorres:2011pr}, i.e., we simply change the integration over momenta by a sum over the discrete values of the momenta allowed by the periodic conditions in the box. We denote the latter one by $\tilde G(s,L)$, and it can be obtained through:
\begin{eqnarray}
\label{eq:difference}
\tilde G(s,L) - G(s) &=&
\lim_{q_{\rm max} \rightarrow \infty} \Big ( {1 \over L^3} \sum_{q_i}^{q_{\rm max}} I(q_i) - \int_{q<q_{\rm max}}{d^3 q \over (2\pi)^3} I(q) \Big ) \, .
\label{gtilde}
\end{eqnarray}
In this equation the discrete momenta in the sum are given by $\vec q = {2 \pi \over L} \vec n ~~ ( \vec n \in \mathcal{Z}^3 )$ and the function $I(q_i)$ is
\begin{eqnarray}
I(q_i) =  {1 \over 2 \omega_1(\vec q) \omega_2(\vec q)} {\omega_1(\vec q) + \omega_2(\vec q) \over E^2 - (\omega_1(\vec q) + \omega_2(\vec q))^2} \, ,
\label{ifun}
\end{eqnarray}
where $\omega_{1,2}(\vec q) = \sqrt{m_{1,2}^2 + \vec q^2}$. We show the real part of $\tilde G(s,L) - G(s)$ in Fig.~\ref{fig:difference} as a function of $q_{\rm max}$, where $L$ is fixed to be 2.5 $m_\pi^{-1}$ and $E$ to be $800$ MeV. Its convergence is good when $q_{\rm max}$ is larger than 3000 MeV. However, we shall still make an average of this quantity for smaller values of $q_{\rm max}$ in order to save the computational time~\cite{Doring:2011jh,MartinezTorres:2011pr}.

\section{The Energy Levels in the Chiral Unitary Approach}
\label{sec:energylevels}

To calculate the energy levels of the $K \pi$ scattering amplitude in $P$-wave, we need to find the poles of the $T(s)$ matrix, which are just solutions of the following equation
\begin{eqnarray}
1 - V(s) \tilde G(s,L) = 0 \, .
\label{eq:EL}
\end{eqnarray}
Here $\tilde G(s,L)$ is defined in the finite volume and can be obtained through Eq.~(\ref{gtilde}). From this equation we can clearly see that the energy levels for $K \pi$ $P$-wave scattering are functions of the cubic box size $L$, as well as the pion mass $m_\pi$. In the following sections we shall study their dependence on these variables. In this section we study the volume dependence and in the next section we shall study the pion mass dependence. We note again that our procedures follow closely the method used in Refs.~\cite{Doring:2011jh,MartinezTorres:2011pr,Chen:2012rp,Doring:2011nd,Xie:2012np,Roca:2012rx}.

\begin{figure}[hbt]
\begin{center}
\scalebox{0.7}{\includegraphics{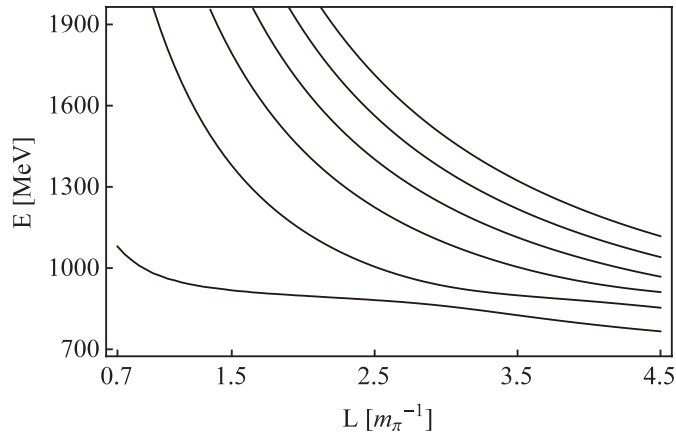}}
\caption{Energy levels as functions of the cubic box size $L$, derived using $\tilde G(s,L)$ from Eq.~(\ref{gtilde}). We perform an average for different $q_{\rm max}$ values between 1200 MeV and 2000 MeV.} 
\label{fig:Level}
\end{center}
\end{figure}

\begin{figure}[hbt]
\begin{center}
\includegraphics[scale=0.8]{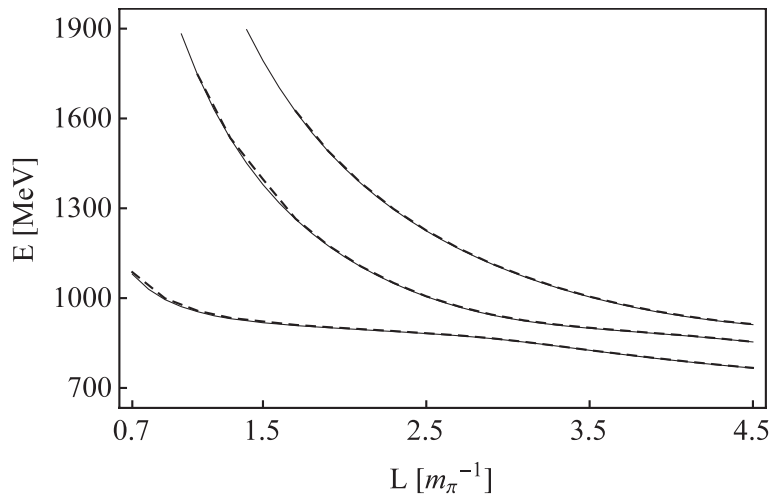}
\caption{Solid curves are $K \pi$ scattering energy levels evaluated using Eq.~(\ref{eq:GDR}) and Eq.~(\ref{gtilde}), and dashed curves are energy levels evaluated using Eq.~(\ref{eq:cutoff}) and Eq.~(\ref{gtilde}), when $q^\prime_{\rm max}=724.70$ MeV~\cite{Xiao:2013mn}.}
\label{fig:cutoff}
\end{center}
\end{figure}

In Fig.~\ref{fig:Level} we show the energy levels as functions of the cubic box size $L$, which are obtained after performing an average for different $q_{max}$ values between 1200 MeV and 2000 MeV. Actually, the results for different $q_{\rm max}$ values are almost the same.
In this figure we have used the dimensional regularization, Eq.~(\ref{eq:GDR}), to calculate Eq.~(\ref{eq:gfunction}) and then calculate $\tilde G(s,L)$ of Eq.~(\ref{gtilde}), while we can also use the cut-off method to calculate Eq.~(\ref{eq:gfunction}):
\begin{eqnarray}
G_{\rm cut off}(s,L)=\int_{q<q^\prime_{\rm max}}{d^3 q \over (2\pi)^3} {1 \over 2 \omega_1(\vec q) \omega_2(\vec q)} {\omega_1(\vec q) + \omega_2(\vec q) \over E^2 - (\omega_1(\vec q) + \omega_2(\vec q))^2}  \, ,
\label{eq:cutoff}
\end{eqnarray}
which can be inserted into Eq.~(\ref{gtilde}) and then calculate $\tilde G(s,L)$. The energy levels can be similarly calculated and the results are shown in Fig.~\ref{fig:cutoff}. We note that the cutoff used in Eq.~(\ref{eq:cutoff}), denoted as $q^\prime_{\rm max}$, is different from $q_{\rm max}$ used in Eq.~(\ref{gtilde}). We choose $q^\prime_{\rm max}$ to be 724.70 MeV following Ref.~\cite{Xiao:2013mn}. It is significantly larger than the discrete momentum $2 \pi / L = 433$ MeV when $L$ is around 2.0 $m_\pi^{-1}$.

\begin{figure}[hbt]
\begin{center}
\scalebox{0.6}{\includegraphics{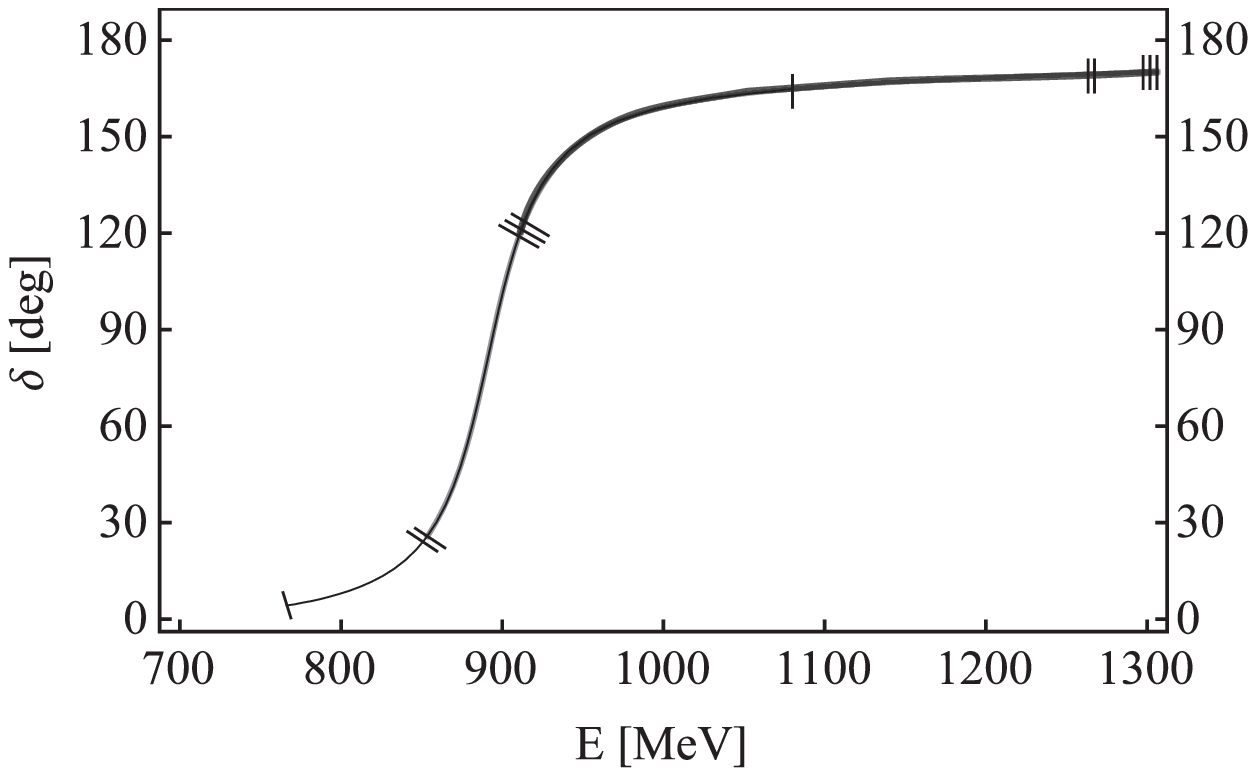}}
\scalebox{0.6}{\includegraphics{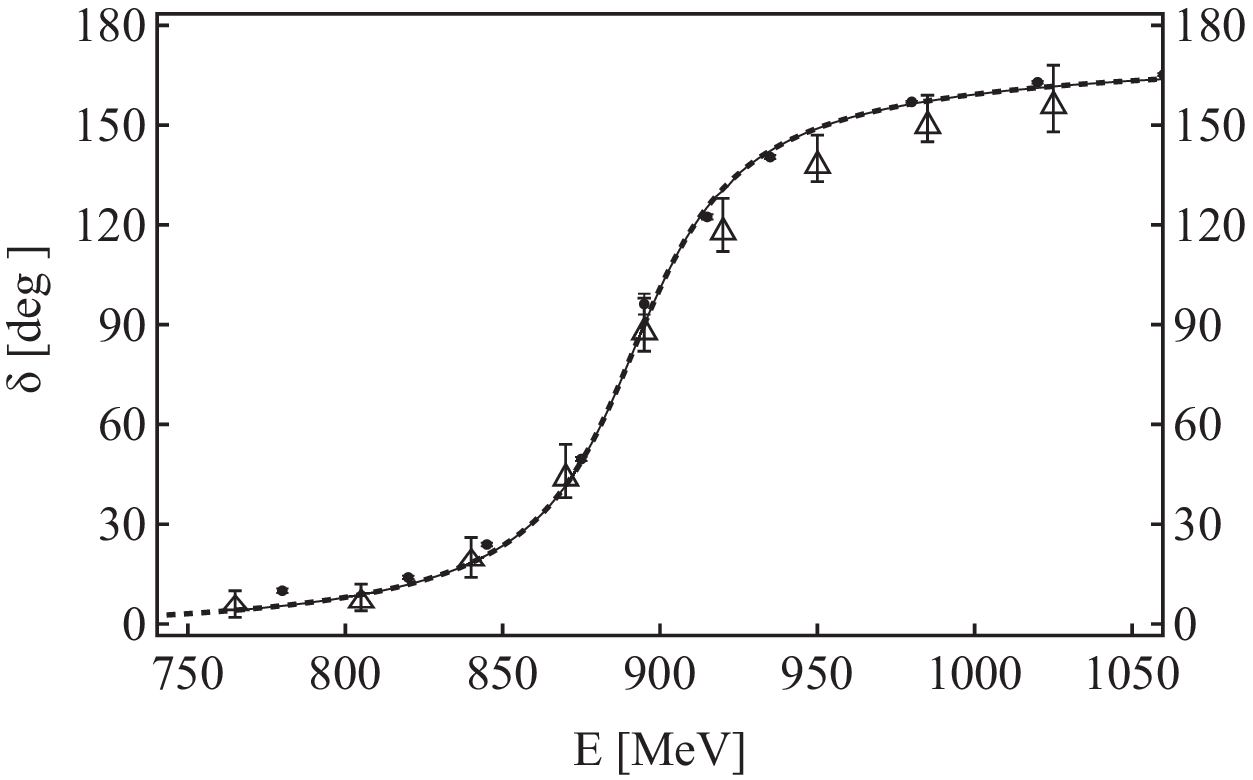}}
\caption{On the left hand side, the curves ended with $|$-$|$, $\|$-$\|$ and $|\|$-$|\|$ correspond to the phase shifts extracted from the first(lowest), the second and the third energy levels of Fig.~\ref{fig:Level}, respctively. On the right hand side, the solid curve is the phase shift extracted from the lowest energy level, the dashed curve is the phase shifts calculated in the infinite volume, and the experimental data are denoted as solid circles~\cite{Estabrooks:1977xe} and triangles~\cite{Mercer:1971kn}.}
 \label{fig:PhaseRead}
\end{center}
\end{figure}

The phase shift can be extracted from these energy levels. To do this we follow the procedure used in Ref.~\cite{Doring:2011vk}, and use Eq.~(\ref{eq:delta}) to calculate the $K \pi$ $P$-wave phase shifts, where the scattering amplitudes $T(E,L)$ are obtained using the energy levels shown in Fig.~(\ref{fig:Level}):
\begin{eqnarray}\label{eq:T11}
T(E, L) &=& { V(E) \over 1 - V(E) G(E) } = { \tilde G(E,L)^{-1} \over 1 - \tilde G(E,L)^{-1} G(E) } \, .
\end{eqnarray}
Here we have used Eq.~(\ref{eq:EL}), i.e., $V(s)^{-1} = \tilde G(s,L)$. Although these procedures can be done for all energy levels, the lowest energy level should be the best one, because we are using the chiral unitary approach which is an effective theory for low energies. Accordingly, we use the lowest energy level to evaluate phase shifts, and the result is shown in Fig.~\ref{fig:PhaseRead}. For comparison, we also show the phase shifts evaluated using the second and the third energy levels.

\begin{figure}[hbt]
\begin{center}
\scalebox{0.62}{\includegraphics{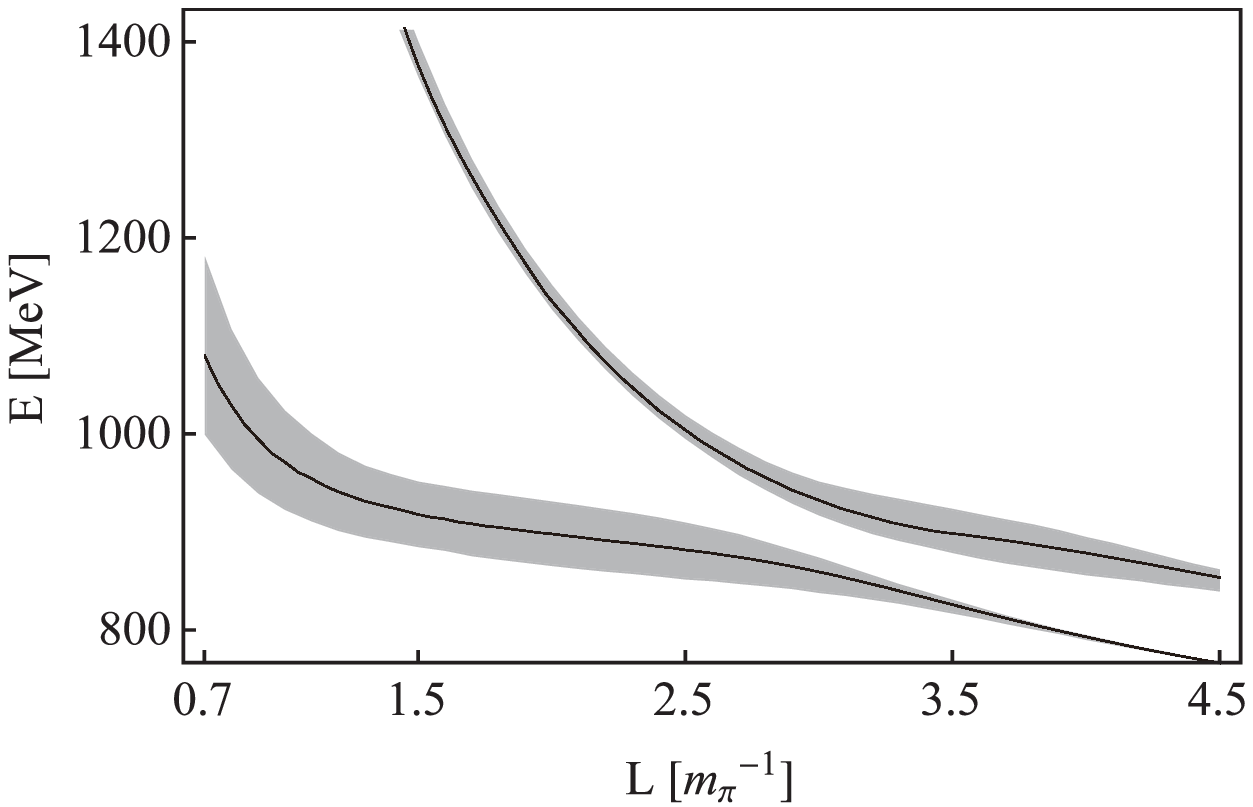}}
\scalebox{0.65}{\includegraphics{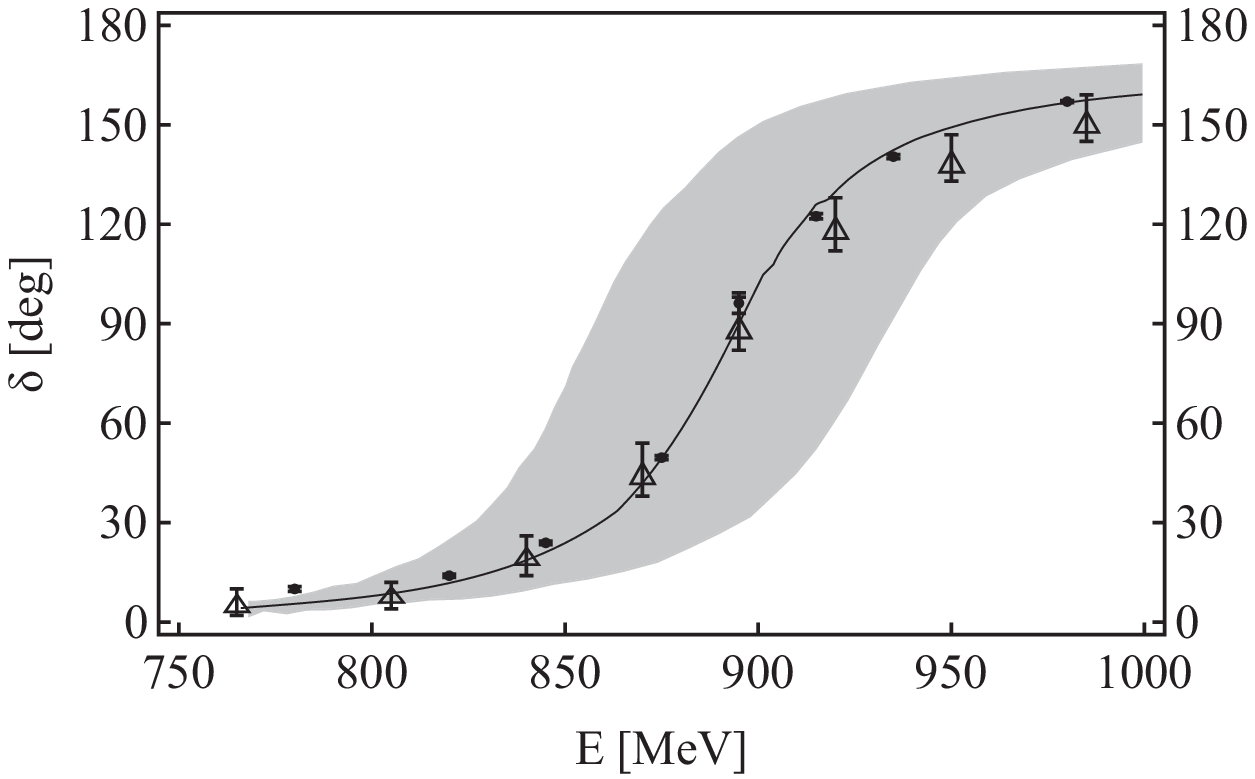}}
\caption{Uncertainties of energy levels and phase shifts.} \label{fig:certainty}
\end{center}
\end{figure}

Using the phase shift $\delta(E)$ we can fit the physical quantities for the $K^*$ meson, and evaluate $m_{K^*}$, $g_{K^* \pi K}$ and $\Gamma_{K^*}$. We note that $m_{K^*}$ is the $K^*$ mass we obtained, i.e., one of our outputs; while $M_{K^*}$ is the bare $K^*$ mass, i.e., one of our inputs. To do that, we use the following two equations in Refs.~\cite{Chen:2012rp,sasa} to extract the $K^*$ properties:
\begin{eqnarray} \label{eq:KstarWidth}
\cot \delta(s) = {m_{K^*}^2 - s \over \sqrt s~\Gamma_{K^*}(s)} \, ,
~~~{\rm and }~~~
\Gamma_{K^*}(s) = {p^3 \over s} {g^2_{K^* \pi K} \over 8 \pi} \, .
\end{eqnarray}
We note that the factor $8 \pi$ in the second equation is our normalization, while in Ref.~\cite{Prelovsek:2013ela} the authors use $6 \pi$. The results from fitting the phase shifts calculated using the lowest $K \pi$ energy level are
\begin{eqnarray}
m_{K^*} = 894.89_{-37.77}^{+39.75} {\rm~MeV}\, , g_{K^* \pi K}=6.48_{-0.12}^{+0.13} \, , \Gamma_{K^*} = 50.68_{-8.00}^{+8.24} {\rm~MeV} \, .
\label{Kstarmass1}
\end{eqnarray}
In these results the theoretical uncertainties are estimated following Ref.~\cite{Chen:2012rp}, where we assume that the uncertainties of the three parameters $G_V$, $M_{K^*}$ and $f$ in Eq.~(\ref{eq:Vmatrix}) are all about 4\%. The uncertainties of the energy levels and phase shifts are shown in Fig.~\ref{fig:certainty}. Particularly, the uncertainty of phase shifts is quite large around $E = 900{\rm~MeV}$. However, the fitted results shown in Eq.~(\ref{Kstarmass1}) have moderate and acceptable uncertainties, suggesting our method is ``stable'' (see also the discussions in Sec.~\ref{sec:luscher}).

Similarly, we can fit the second and the third energy levels. We find that the results do not change much: the results from fitting the phase shifts calculated using both the first and the second energy levels are (overlapped points are counted just once):
\begin{eqnarray}
m_{K^*} = 894.78{\rm~MeV}\, , g_{K^* \pi K}=6.34 \, , \Gamma_{K^*} = 48.49 {\rm~MeV} \, ,
\end{eqnarray}
and the results from fitting the phase shifts calculated using all the three energy levels are (overlapped points are counted just once):
\begin{eqnarray}
m_{K^*} = 894.84{\rm~MeV}\, , g_{K^* \pi K}=6.31 \, , \Gamma_{K^*} = 48.04 {\rm~MeV} \, .
\end{eqnarray}

\section{Dependence on the Pion Mass}
\label{sec:pimass}

As we know, due to the computer limitation, the Lattice QCD calculations usually use a non-physical pion mass. Therefore, in this section, we also use non-physical pion masses to study the mass and decay width of the $K^*$ meson, in order to compare with the Lattice QCD result. We define $m_\pi^0$ to be the physical $\pi$ mass, and now $m_\pi$ is a free parameter. We change it from $m_\pi^0$ to $3 m_\pi^0$. At the same time other parameters can also change with $m_\pi$. We follow the same approach of Refs.~\cite{Chen:2012rp,Boucaud:2007uk,Beane:2007xs,Noaki:2008gx,Pelaez:2010fj}, where the variation of $f$ as a function of $m_\pi$ is
\begin{equation}
\frac{f(m_{\pi})}{f(m^0_{\pi})}= 1 +0.048((\frac{m_{\pi}}{m^0_{\pi}})^2-1),
\end{equation}
with $f(m^0_{\pi})=86.22$~MeV. The coupling $G_V$ is related to $f$ \cite{sakurai,Bando:1987br,Ecker:1989yg,Zhou:2014ila}, as $G_V = f / \sqrt 2$, valid to leading order, consequently, we keep $G_V/f$ unchanged. The kaon mass $m_K$ can also change with the pion mass $m_\pi$, and we use the following relation~\cite{Ren:2012aj}:
\begin{equation}
m_K^2 = a + bm_\pi ^2,
\label{eq.mK}
\end{equation}
where $a = 0.29{\rm~GeV}^2$, and $b = 0.67$. We note that the Lattice calculations also use non-physical kaon masses~\cite{Fu:2012tj,Lang:2012sv}, but all these values are not much different from the physical one. Accordingly, in this paper we shall first keep it unchanged, then use the kaon mass in Eq.~(\ref{eq.mK}), and finally use the same values of $m_K$ as the Lattice ones~\cite{Fu:2012tj,Lang:2012sv} in order to compare our results with theirs. On the other hand, the bare $K^*$ mass, $M_{K^*}$ in Eq.~(\ref{eq:Vmatrix}), provides the link of the theory to a genuine component of the $K^*$ meson, not related to the $K \pi$ component, and we assume it to be  $m_{\pi}$ independent.

\begin{figure}[hbt]
\begin{center}
\scalebox{0.45}{\includegraphics{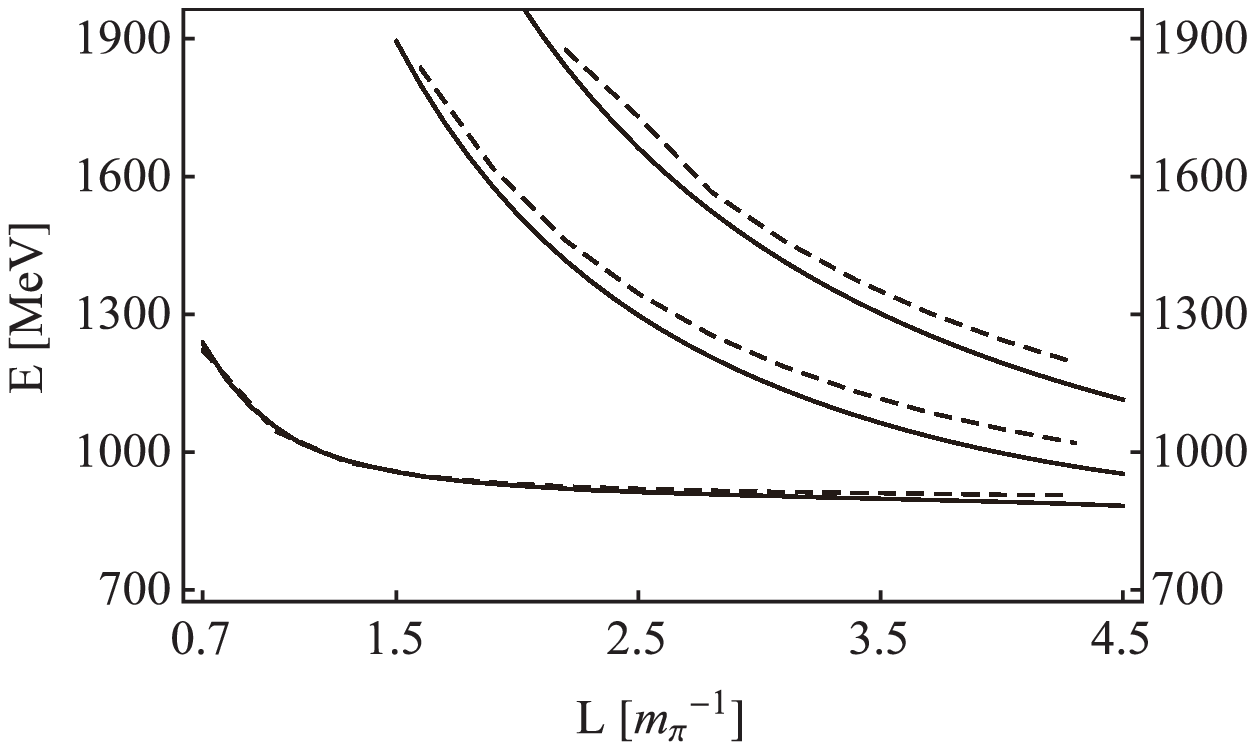}}
\scalebox{0.45}{\includegraphics{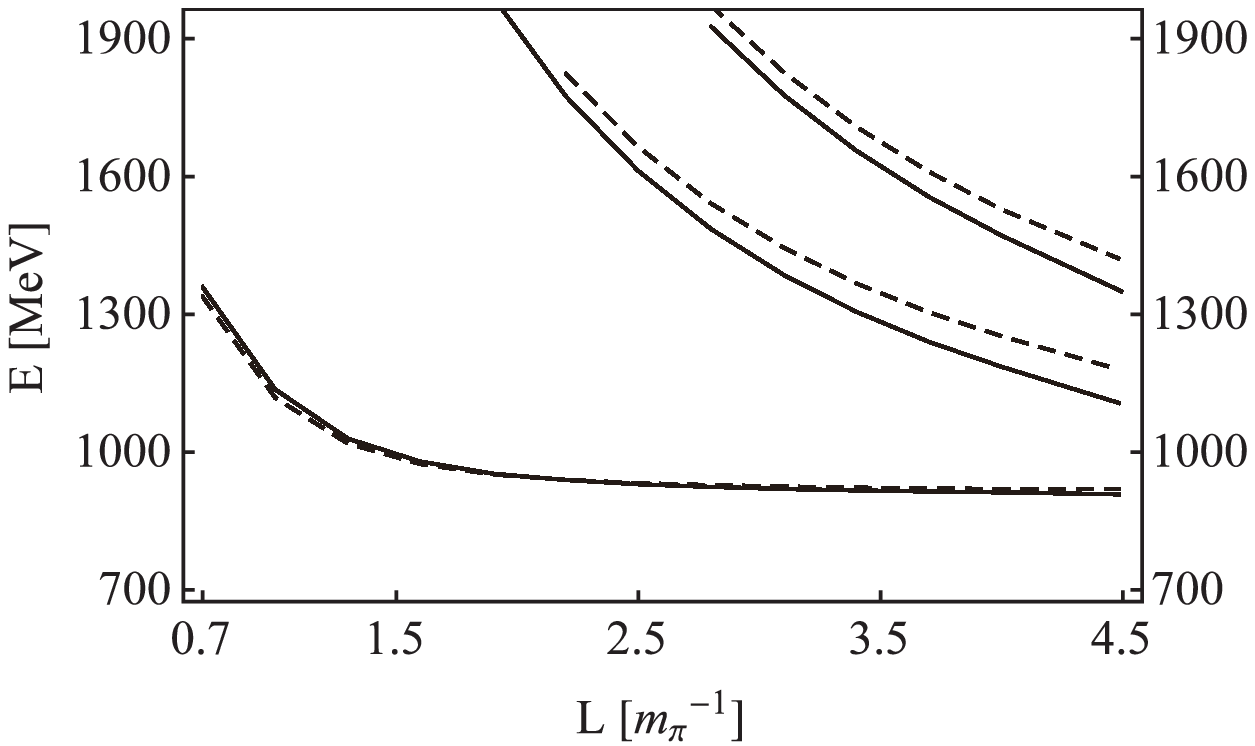}}
\scalebox{0.45}{\includegraphics{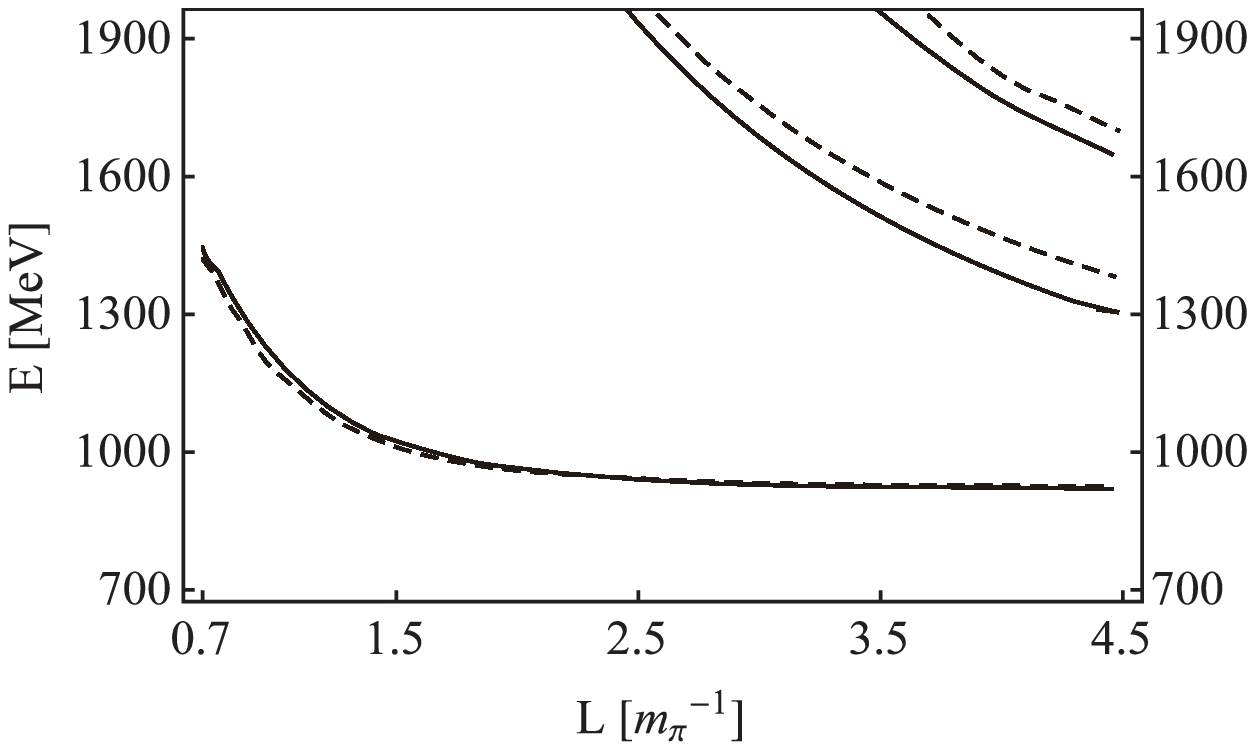}}
\caption{Energy levels as functions of the cubic box size $L$. The left, middle and right figures correspond to $m_\pi = 1.5~m_\pi^0$, $m_\pi = 2.0~m_\pi^0$ and $m_\pi = 2.5~m_\pi^0$, respectively. The solid curves are obtained using the physical kaon mass $m_K = 496 {\rm~MeV}$, and the dotted curves are obtained using the non-physical kaon mass evaluated using Eq.~(\ref{eq.mK}).} \label{fig:mpi}
\end{center}
\end{figure}

To calculate the energy levels we follow the same procedures which have been used in the previous section. The result is shown in Fig.~\ref{fig:mpi} where we have used $m_\pi = 1.5~m_\pi^0$ (left), $m_\pi = 2.0~m_\pi^0$ (middle) and $m_\pi = 2.5~m_\pi^0$ (right). The solid curves are obtained using the physical kaon mass $m_K = 496 {\rm~MeV}$, and the dotted curves are obtained using the non-physical kaon mass evaluated using Eq.~(\ref{eq.mK}). We can see that the results obtained using these different kaon masses do not differ much. Here, we note that the $x$-coordinate is expressed in units of $m_\pi^{-1}$, not $(m_\pi^0)^{-1}$.
\begin{figure}[hbt]
\begin{center}
\scalebox{0.45}{\includegraphics{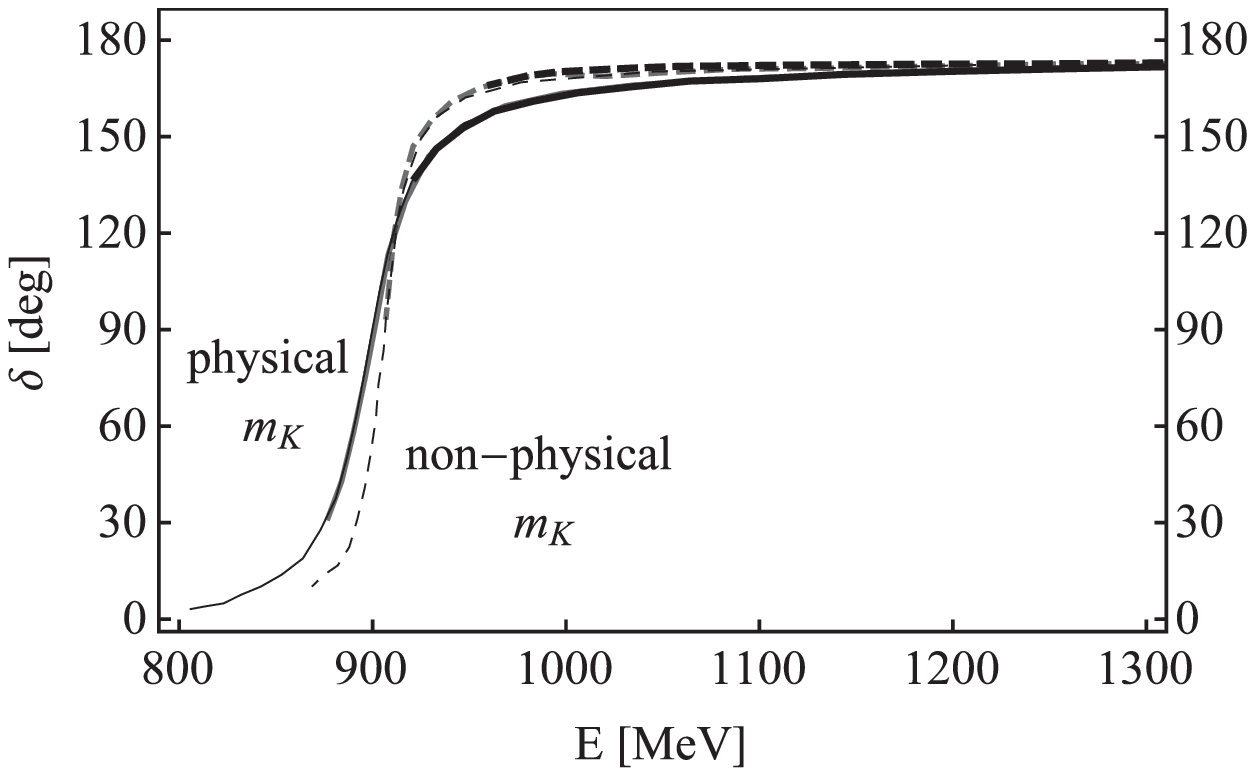}}
\scalebox{0.45}{\includegraphics{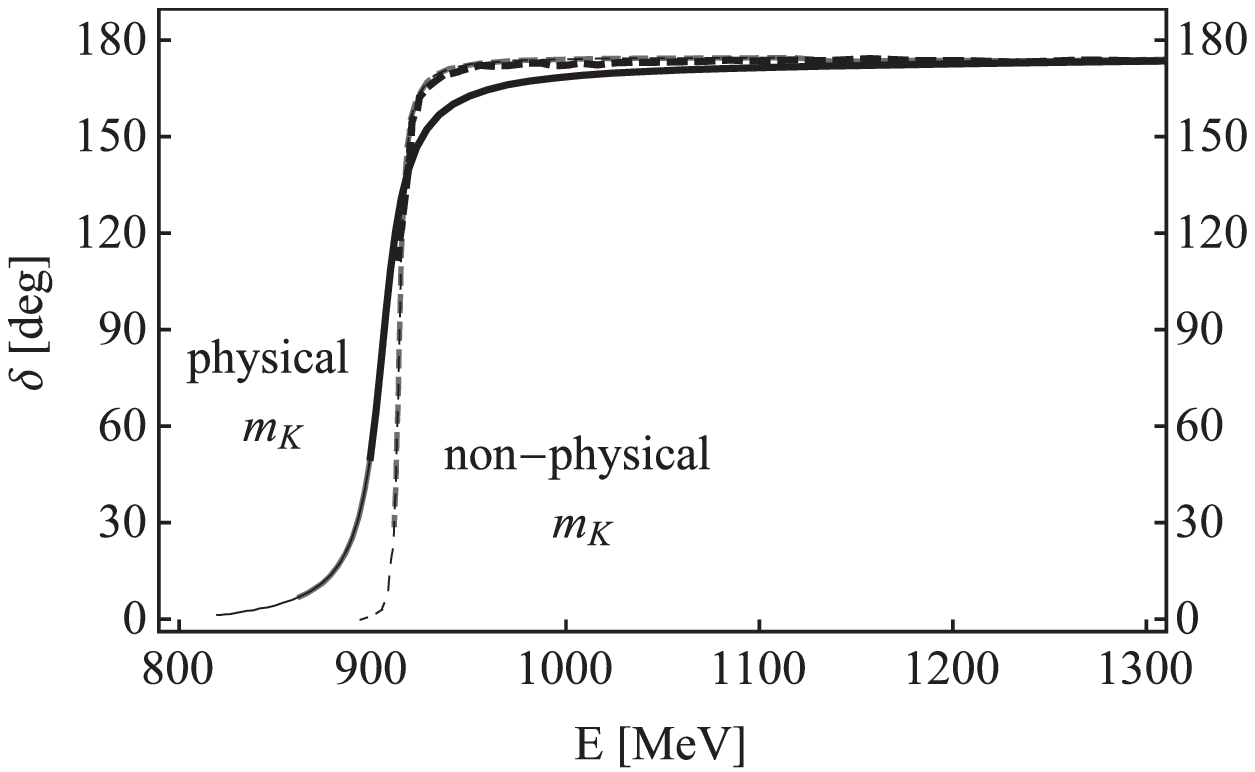}}
\scalebox{0.45}{\includegraphics{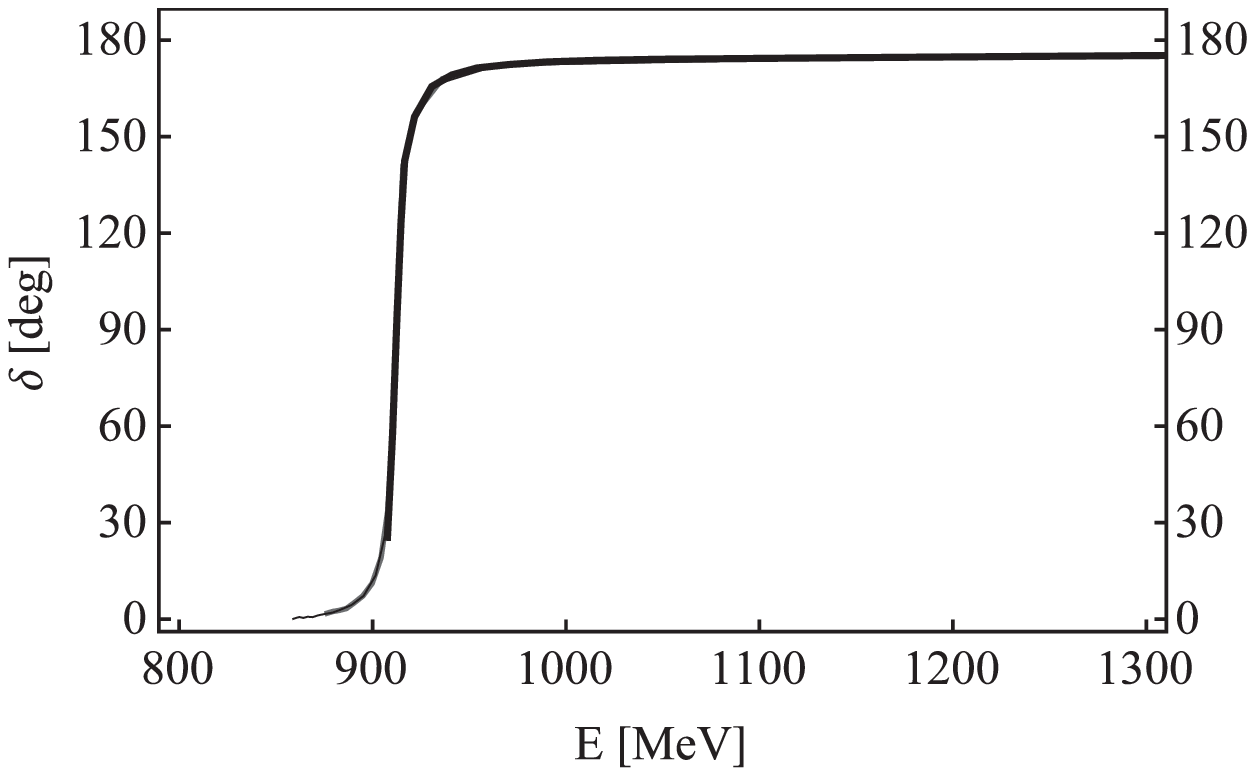}}
\caption{$K \pi$ phase shifts with different pion masses. The left, middle and right figures correspond to $m_\pi = 1.5~m_\pi^0$, $m_\pi = 2.0~m_\pi^0$ and $m_\pi = 2.5~m_\pi^0$, respectively. The solid curves are obtained using the physical kaon mass $m_K = 496 {\rm~MeV}$, and the dotted curves are obtained using the non-physical kaon mass evaluated using Eq.~(\ref{eq.mK}).} \label{fig:mpips}
\end{center}
\end{figure}

\begin{figure}[hbt]
\begin{center}
\scalebox{0.6}{\includegraphics{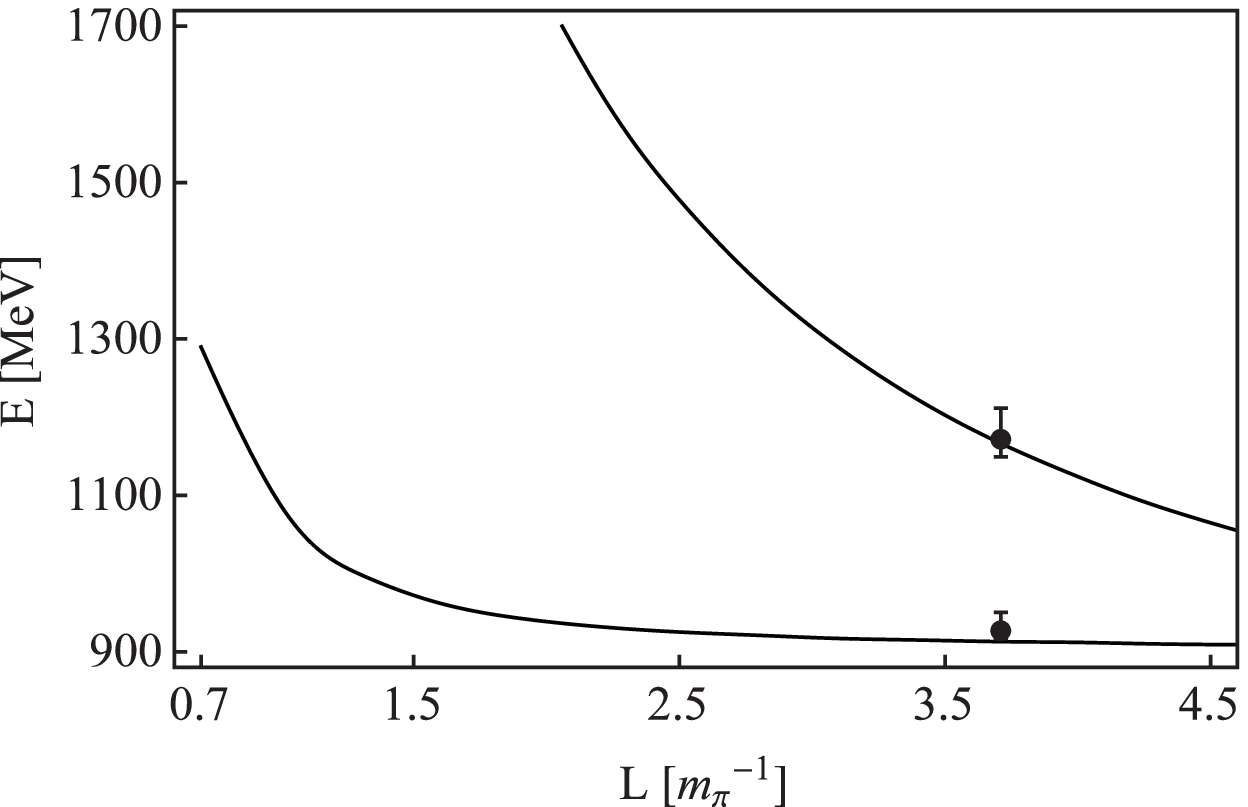}}
\scalebox{0.6}{\includegraphics{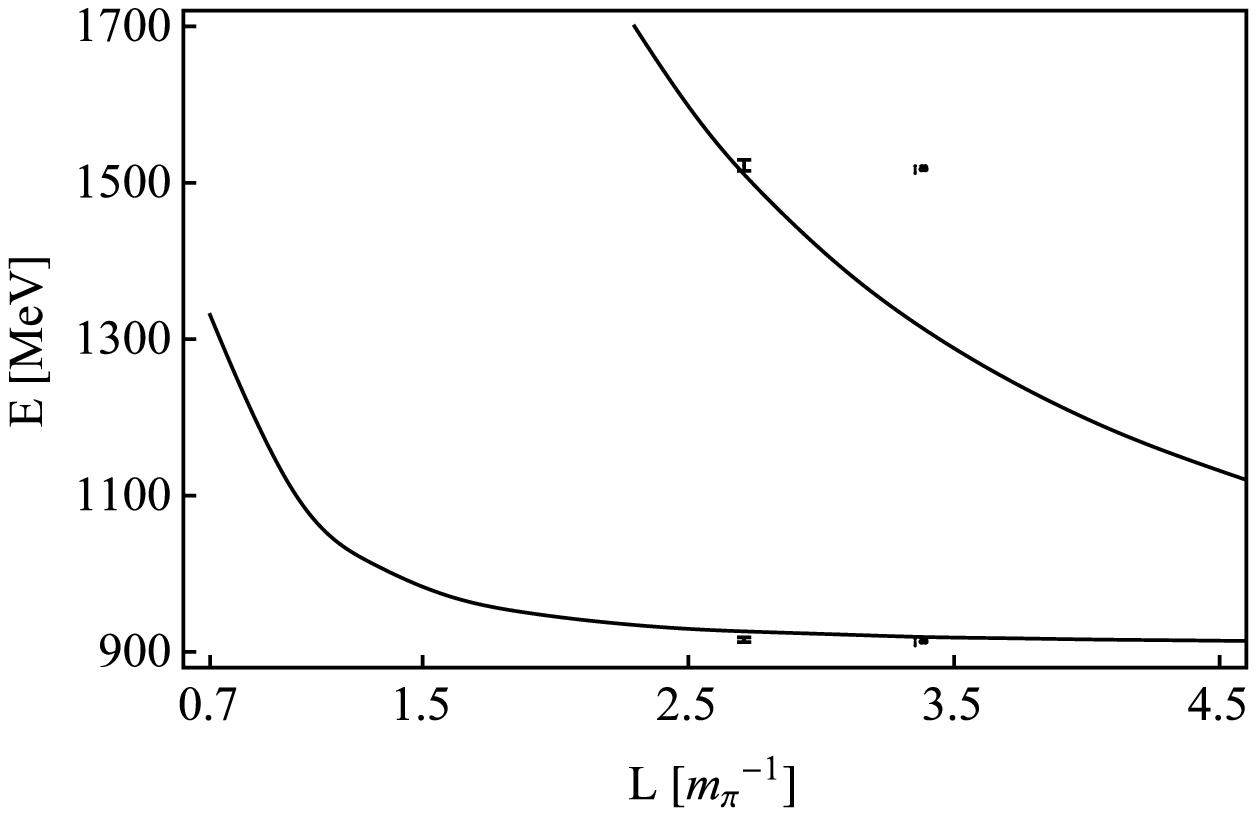}}
\caption{The $K \pi$ energy levels obtained using the chiral unitary theory. In the left panel we use $m_\pi =240$ MeV and
$m_K=548$ MeV to compare with the results of Ref.~\cite{Fu:2012tj}, and in the right panel we use $m_\pi=266$ MeV and $m_K=552$ MeV to compare with the results of Ref.~\cite{Lang:2012sv}. We also show the Lattice results for comparisons~\cite{Fu:2012tj,Lang:2012sv}. Here we do not evaluate their uncertainties, but note that they are similar to Fig.~\ref{fig:certainty}.}
\label{fig:Lattice}
\end{center}
\end{figure}
These energy levels can be used to calculate the phase shifts again following our previous procedures. The results are shown in Fig.~\ref{fig:mpips}, where again we have used $m_\pi = 1.5~m_\pi^0$ (left), $m_\pi = 2.0~m_\pi^0$ (middle) and $m_\pi = 2.5~m_\pi^0$ (right). The solid curves are obtained using the physical kaon mass $m_K = 496 {\rm~MeV}$, and the dashed curves are obtained using the non-physical kaon mass evaluated using Eq.~(\ref{eq.mK}). We note that the dashed curve on the right of Fig.~\ref{fig:mpips} vanishes, because the sum of $2.5~m_\pi^0$ and non-physical kaon mass $m_K = 610{\rm~MeV}$ is already above the $K^*$ threshold.

Now we can compare our results with the Lattice results of Refs.~\cite{Fu:2012tj,Lang:2012sv}, where $m_\pi =240$ MeV and
$m_K=548$ MeV are used in Ref.~\cite{Fu:2012tj}, and $m_\pi=266$ MeV and $m_K=552$ MeV are used in Ref.~\cite{Lang:2012sv}. We show their comparisons in Fig.~\ref{fig:Lattice} and Tables~\ref{tab:comparedata1} and ~\ref{tab:comparedata2}, where $E_1$ and $E_2$ are on the lowest and the second energy levels, and $\delta_1$ and $\delta_2$ are extracted from $E_1$ and $E_2$, respectively. We find that the energy levels and the extracted phase shifts are similar, and so our results compare favorably with those lattice results obtained in Refs.~\cite{Fu:2012tj,Lang:2012sv}. Again the theoretical errors are obtained by assuming that the uncertainties of the three parameters $G_V$, $M_{K^*}$ and $f$ in Eq.~(\ref{eq:Vmatrix}) are about 4\%. We also show more points in Table~\ref{tab:diffmpi} which may be useful.
\begin{table}[!hbt]
\caption{Comparison with Ref.~\cite{Fu:2012tj}, where $m_\pi = 240$ MeV, $m_K = 548$ MeV and $L = 3$ fm.}
\begin{center}
\label{tab:comparedata1}
\begin{tabular}{c|cc}
\hline \hline
\ & \mbox{$E_1$} & \mbox{$E_2$}
\\ \hline
Our Results & $912.6_{-33.5}^{+33.4} {\rm MeV}$ & $1166.7_{-5.1}^{+5.2}{\rm MeV}$
\\ \hline
Lattice Results & $926.9_{-10.0}^{+23.5} {\rm MeV}$ & $1171.7_{-22.5}^{+40.0} {\rm MeV}$
\\ \hline \hline
\end{tabular}
\end{center}
\end{table}
\begin{table}[!hbt]
\caption{Comparison with Ref.~\cite{Lang:2012sv}, where $m_\pi = 266$ MeV, $m_K = 552$ MeV and $L = 1.98$ fm.}
\begin{center}
\label{tab:comparedata2}
\begin{tabular}{c|cccc}
\hline \hline
\ & \mbox{$E_1$} & \mbox{$E_2$} & \mbox{$\delta_1$} & \mbox{$\delta_2$}
\\ \hline
Our Results & $926.2_{-36.8}^{+36.0}{\rm MeV}$ & $1511.4_{-7.5}^{+9.6} {\rm MeV}$ & $158.05^\circ$ $_{-8.45^\circ}^{+8.52^\circ}$ & $175.52^\circ$ $_{-3.62^\circ}^{+2.79^\circ}$
\\ \hline
Lattice Results & $915.6 \pm 3.0 {\rm MeV}$ & $1522.3 \pm 7.0 {\rm MeV}$ & $160.61^\circ \pm 0.73 ^\circ$ & $177.0^\circ \pm 2.6^\circ$
\\ \hline  \hline
\end{tabular}
\end{center}
\end{table}

\begin{table}[hbt]
\begin{center}
\caption{Some examples of energy levels and phase shifts.}
\begin{tabular}{c|c|c|c|c|c|c|c|c}
\hline \hline
\cline{2-9}\ & \multicolumn{2}{c|} {} & \multicolumn{2}{c|} {$m_K=500{\rm MeV}$} & \multicolumn{4}{c} {$m_K=600{\rm MeV}$}
\\ \hline
\multirow{8}{*} {$m_\pi = 250 {\rm MeV}$} & \multicolumn{2}{c|} {\multirow{2}{*}{$L=1.5 fm$}} & \multicolumn{2}{c|} {$E_1=944.7 {\rm MeV}, \delta_1=158.41^\circ$} & \multicolumn{4}{c} {$E_1=944.1 {\rm MeV}, \delta_1=170.64^\circ$}
\\  & \multicolumn{2}{c|}{} & \multicolumn{2}{c|} {$E_2=1836.0 {\rm MeV}, \delta_2=178.45^\circ$} & \multicolumn{4}{c} {$E_2=1889.0 {\rm MeV}, \delta_2=179.02^\circ$}
\\ \cline{2-9} \ & \multicolumn{2}{c|}{\multirow{2}{*}{$L=2.0 fm$}} & \multicolumn{2}{c|} {$E_1=923.9 {\rm MeV}, \delta_1=145.20^\circ$} & \multicolumn{4}{c} {$E_1=927.8 {\rm MeV}, \delta_1=165.35^\circ$}
\\  & \multicolumn{2}{c|}{} & \multicolumn{2}{c|} {$E_2=1477.6 {\rm MeV}, \delta_2=174.33^\circ$} & \multicolumn{4}{c} {$E_2=1540.0 {\rm MeV}, \delta_2=175.78^\circ$}
\\ \cline{2-9} \ & \multicolumn{2}{c|}{\multirow{2}{*} {$L=2.5 fm$}} & \multicolumn{2}{c|} {$E_1=913.8 {\rm MeV}, \delta_1=127.68^\circ$} & \multicolumn{4}{c} {$E_1=921.2 {\rm MeV}, \delta_1=157.0^\circ$}
\\  & \multicolumn{2}{c|}{} & \multicolumn{2}{c|} {$E_2=1271.8 {\rm MeV}, \delta_2=172.42^\circ$} & \multicolumn{4}{c} {$E_2=1342.0 {\rm MeV}, \delta_2=174.32^\circ$}
\\ \cline{2-9} \ & \multicolumn{2}{c|} {\multirow{2}{*}{$L=3.0 fm$}} & \multicolumn{2}{c|} {$E_1=908.4 {\rm MeV}, \delta_1=109.44^\circ$} & \multicolumn{4}{c} {$E_1=918.1 {\rm MeV}, \delta_1=148.59^\circ$}
\\  & \multicolumn{2}{c|}{} & \multicolumn{2}{c|} {$E_2=1143.0 {\rm MeV}, \delta_2=170.96^\circ$} & \multicolumn{4}{c} {$E_2=1218.0 {\rm MeV}, \delta_2=173.86^\circ$}
\\ \hline \hline

\multirow{8}{*}{$m_\pi = 300 {\rm MeV}$} & \multicolumn{2}{c|} {\multirow{2}{*}{$L=1.5 fm$}} & \multicolumn{2}{c|} {$E_1=940.6 {\rm MeV}, \delta_1=163.41^\circ$} & \multicolumn{4}{c} {$E_1=939.9 {\rm MeV}$}
\\  & \multicolumn{2}{c|}{} & \multicolumn{2}{c|} {$E_2=1844.0 {\rm MeV}, \delta_2=178.76^\circ$} & \multicolumn{4}{c} {$E_2=1897.0 {\rm MeV}$}
\\ \cline{2-9} \ & \multicolumn{2}{c|} {\multirow{2}{*}{$L=2.0 fm$}} & \multicolumn{2}{c|} {$E_1=924.6 {\rm MeV}, \delta_1=153.26^\circ$} & \multicolumn{4}{c} {$E_1=927.5 {\rm MeV}$}
\\  & \multicolumn{2}{c|}{} & \multicolumn{2}{c|} {$E_2=1494.2 {\rm MeV}, \delta_2=175.20^\circ$} & \multicolumn{4}{c} {$E_2=1557.1 {\rm MeV}$}
\\ \cline{2-9} \ & \multicolumn{2}{c|} {\multirow{2}{*}{$L=2.5 fm$}} & \multicolumn{2}{c|} {$E_1=917.3 {\rm MeV}, \delta_1=139.31^\circ$} & \multicolumn{4}{c} {$E_1=922.8 {\rm MeV}$}
\\ & \multicolumn{2}{c|}{} & \multicolumn{2}{c|} {$E_2=1293.9 {\rm MeV}, \delta_2=174.03^\circ$} & \multicolumn{4}{c} {$E_2=1364.4 {\rm MeV}$}
\\ \cline{2-9} \ & \multicolumn{2}{c|} {\multirow{2}{*}{$L=3.0 fm$}} & \multicolumn{2}{c|} {$E_1=913.7 {\rm MeV}, \delta_1=125.50^\circ$} & \multicolumn{4}{c} {$E_1=920.8 {\rm MeV}$}
\\  & \multicolumn{2}{c|}{} & \multicolumn{2}{c|} {$E_2=1168.8 {\rm MeV}, \delta_2=173.07^\circ$} & \multicolumn{4}{c} {$E_2=1244.7 {\rm MeV}$}
\\ \hline \hline

\multirow{8}{*}{$m_\pi = 350 {\rm MeV}$} & \multicolumn{2}{c|} {\multirow{2}{*}{$L=1.5 fm$}} & \multicolumn{2}{c|} {$E_1=937.7 {\rm MeV}, \delta_1=169.80^\circ$} & \multicolumn{4}{c} {$E_1=936.9 {\rm MeV}$}
\\  & \multicolumn{2}{c|}{} & \multicolumn{2}{c|} {$E_2=1867.2 {\rm MeV}, \delta_2=178.96^\circ$} & \multicolumn{4}{c} {$E_2=1920.5 {\rm MeV}$}
\\ \cline{2-9} \ & \multicolumn{2}{c|} {\multirow{2}{*}{$L=2.0 fm$}} & \multicolumn{2}{c|} {$E_1=925.2 {\rm MeV}, \delta_1=163.42^\circ$} & \multicolumn{4}{c} {$E_1=927.1 {\rm MeV}$}
\\  & \multicolumn{2}{c|}{} & \multicolumn{2}{c|} {$E_2=1517.1 {\rm MeV}, \delta_2=176.13^\circ$} & \multicolumn{4}{c} {$E_2=1580.7 {\rm MeV}$}
\\ \cline{2-9} \ & \multicolumn{2}{c|} {\multirow{2}{*}{$L=2.5 fm$}} & \multicolumn{2}{c|} {$E_1=919.9 {\rm MeV}, \delta_1=154.35^\circ$} & \multicolumn{4}{c} {$E_1=923.9 {\rm MeV}$}
\\  & \multicolumn{2}{c|}{} & \multicolumn{2}{c|} {$E_2=1319.3 {\rm MeV}, \delta_2=175.07^\circ$} & \multicolumn{4}{c} {$E_2=1390.7 {\rm MeV}$}
\\ \cline{2-9} \ & \multicolumn{2}{c|} {\multirow{2}{*}{$L=3.0 fm$}} & \multicolumn{2}{c|} {$E_1=917.5 {\rm MeV}, \delta_1=145.01^\circ$} & \multicolumn{4}{c} {$E_1=922.4{\rm MeV}$}
\\  & \multicolumn{2}{c|}{} & \multicolumn{2}{c|} {$E_2=1196.9 {\rm MeV}, \delta_2=174.92^\circ$} & \multicolumn{4}{c} {$E_2=1274.0 {\rm MeV}$}
\\ \hline \hline
\end{tabular}
\label{tab:diffmpi}
\end{center}
\end{table}

\begin{figure}[hbt]
\begin{center}
\scalebox{0.47}{\includegraphics{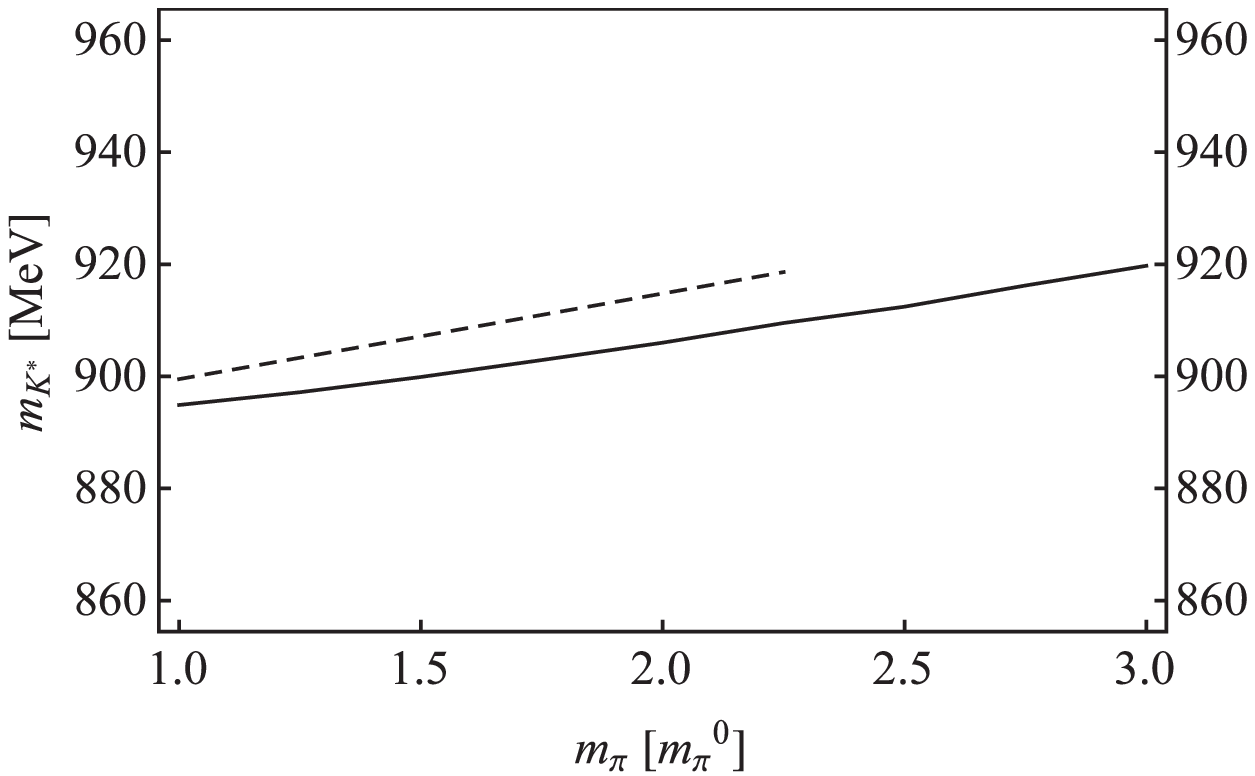}}
\scalebox{0.45}{\includegraphics{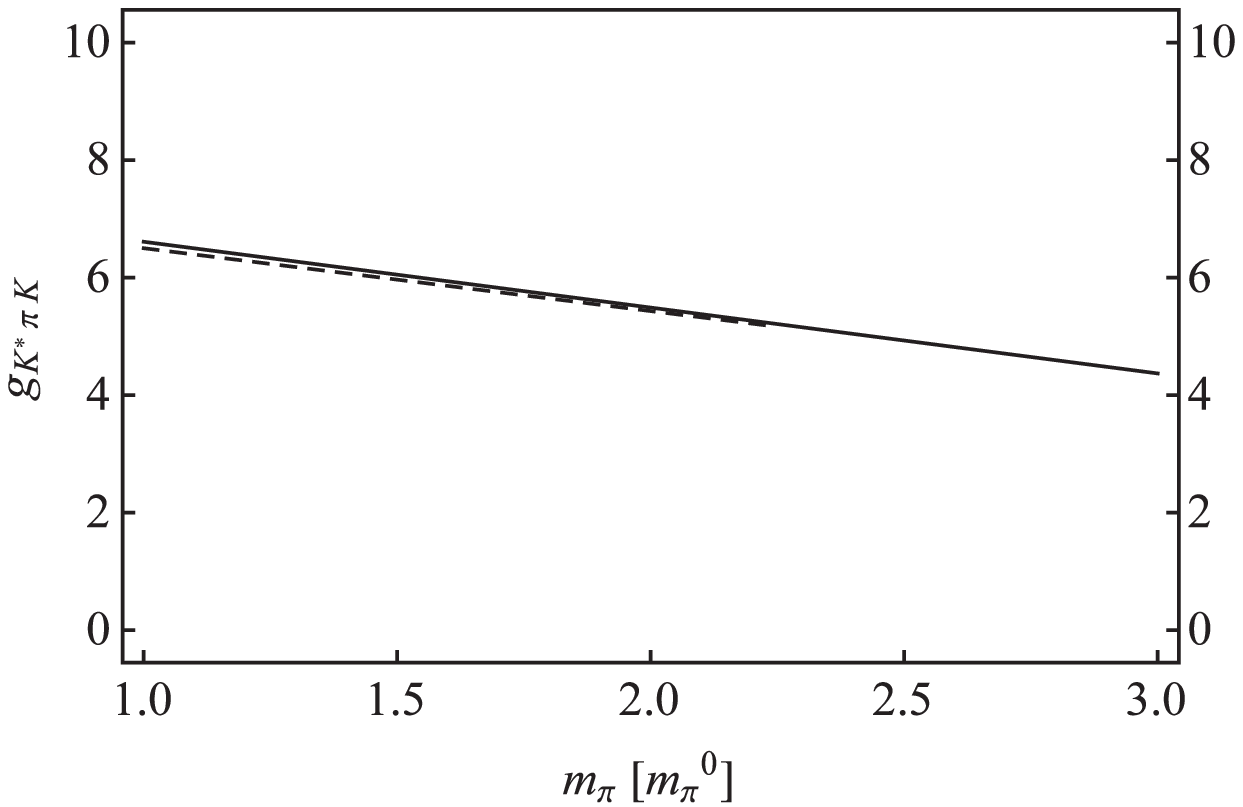}}
\scalebox{0.45}{\includegraphics{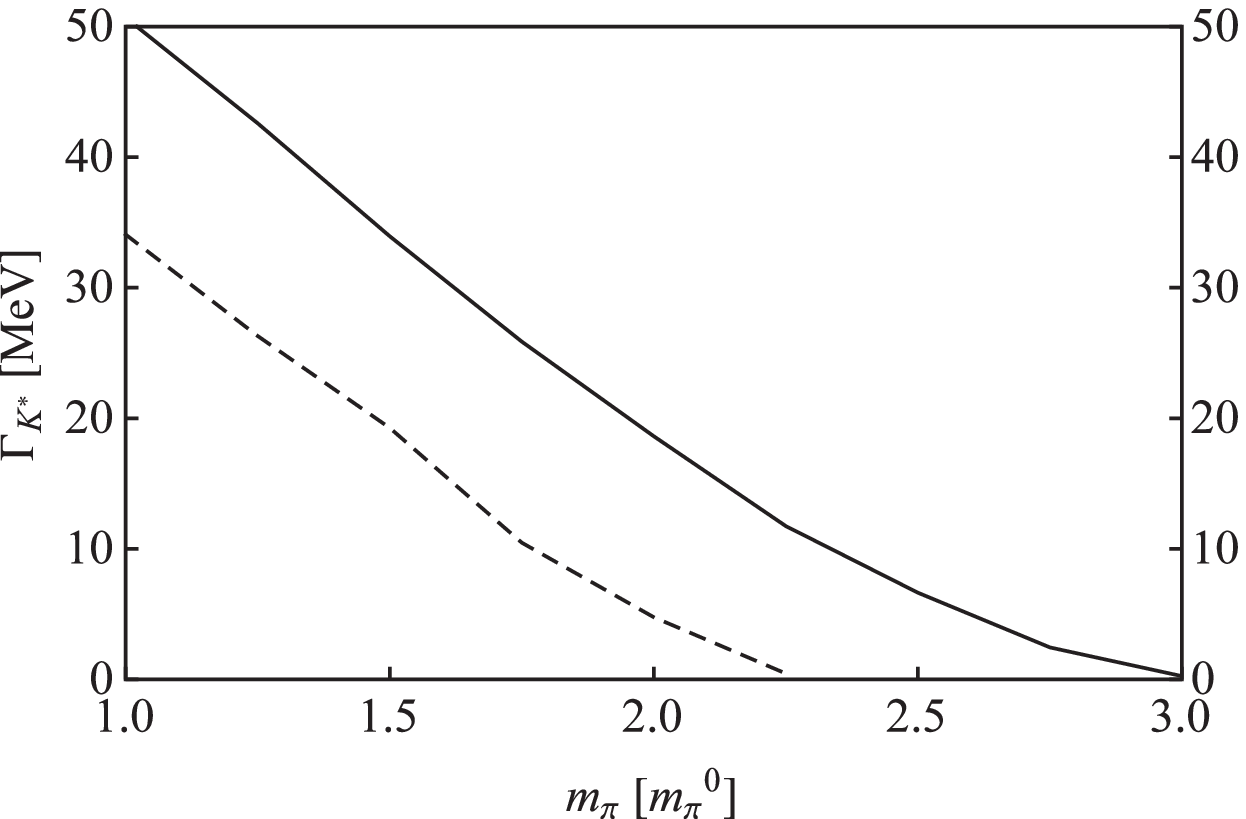}}
\caption{The $K^*$ mass (left), the coupling constant $g_{K^* \pi K}$ (middle) and the decay width $\Gamma_{K^*}$ (right) as functions of $m_\pi$. The solid curves are obtained using the physical kaon mass $m_K = 496 {\rm~MeV}$, and the dashed curves are obtained using the non-physical kaon mass evaluated using Eq.~(\ref{eq.mK}).} \label{fig:Kstarmass}
\end{center}
\end{figure}

Finally, we use Eq.(\ref{eq:KstarWidth}) to fit the phase shifts obtained using the lowest $K \pi$ energy level, and obtain the $K^*$ mass (left), the coupling constant $g_{K^* \pi K}$ (middle) and the decay width $\Gamma_{K^*}$ (right), which are shown in Fig.~\ref{fig:Kstarmass} as functions of $m_\pi$. We can see that the results of $g_{K^* \pi K}$ obtained using the physical kaon mass and non-physical kaon masses in Eq.~(\ref{eq.mK}) are very similar, while the results of $m_{K^*}$ and $\Gamma_{K^*}$ are not so similar. This may be because the phase spaces differ much, although the kaon masses do not differ much. We also note that when using Eq.~(\ref{eq.mK}), the physical kaon mass $m_K = 496{\rm~MeV}$ can not be reached at the physical pion mass $m_\pi = 138{\rm~MeV}$. Therefore, the dashed curves and the solid curves are not connected. Again we can compare our results with the lattice results in Ref.~\cite{Prelovsek:2013ela}, where $m_\pi = 266 {\rm MeV}$, $m_K = 552 {\rm MeV}$ and $L = 1.98 {\rm fm}$. Their results are $m_{K^*} = 891 \pm 14$ MeV and $g_{K^* \pi K} = 5.7 \pm 1.6$, which change to $m_{K^*} = 891 \pm 14$ MeV and $g_{K^* \pi K} = 6.6 \pm 1.9$ in our normalization after taking into account the factor ${8 \pi \over 6 \pi}$. These results are in agreement, within uncertainties, with our result $m_{K^*} = 910.5_{-33.8}^{+34.3}$ MeV and $g_{K^* \pi K} = 5.61_{-0.27}^{+0.21}$.

\begin{figure}[hbt]
\begin{center}
\scalebox{0.47}{\includegraphics{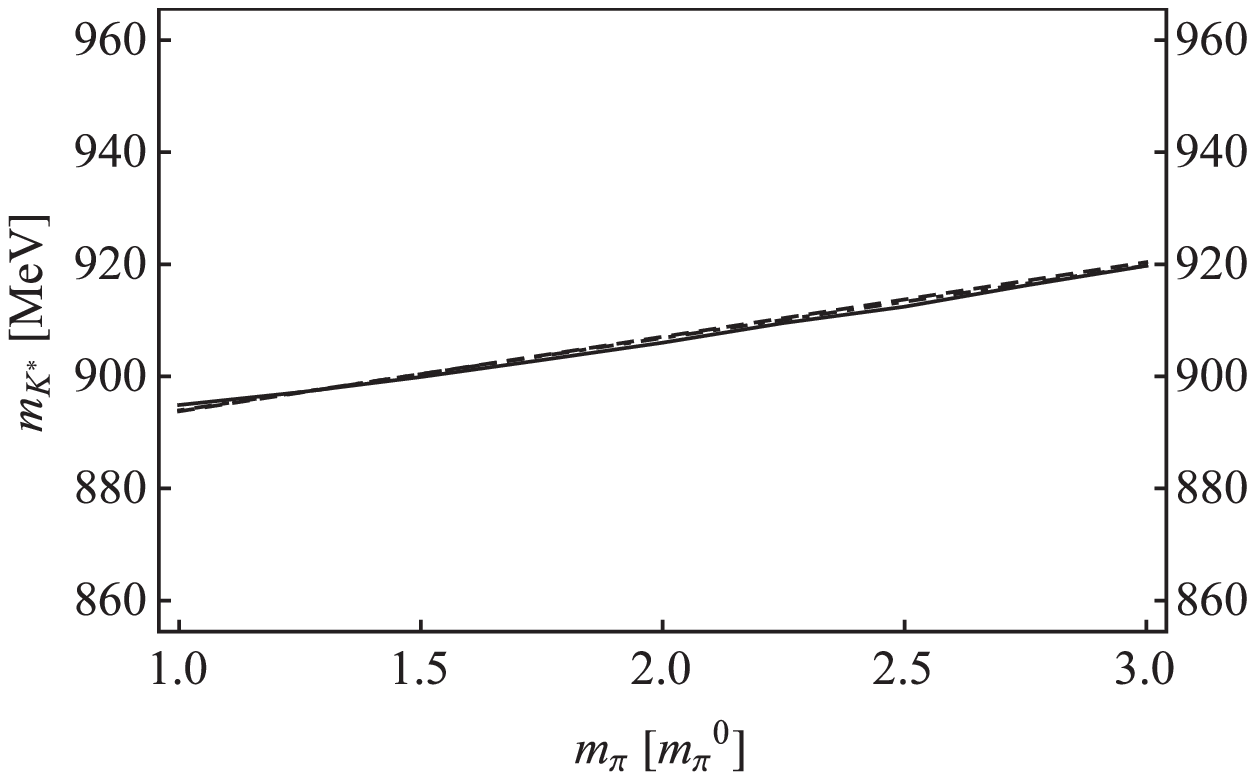}}
\scalebox{0.45}{\includegraphics{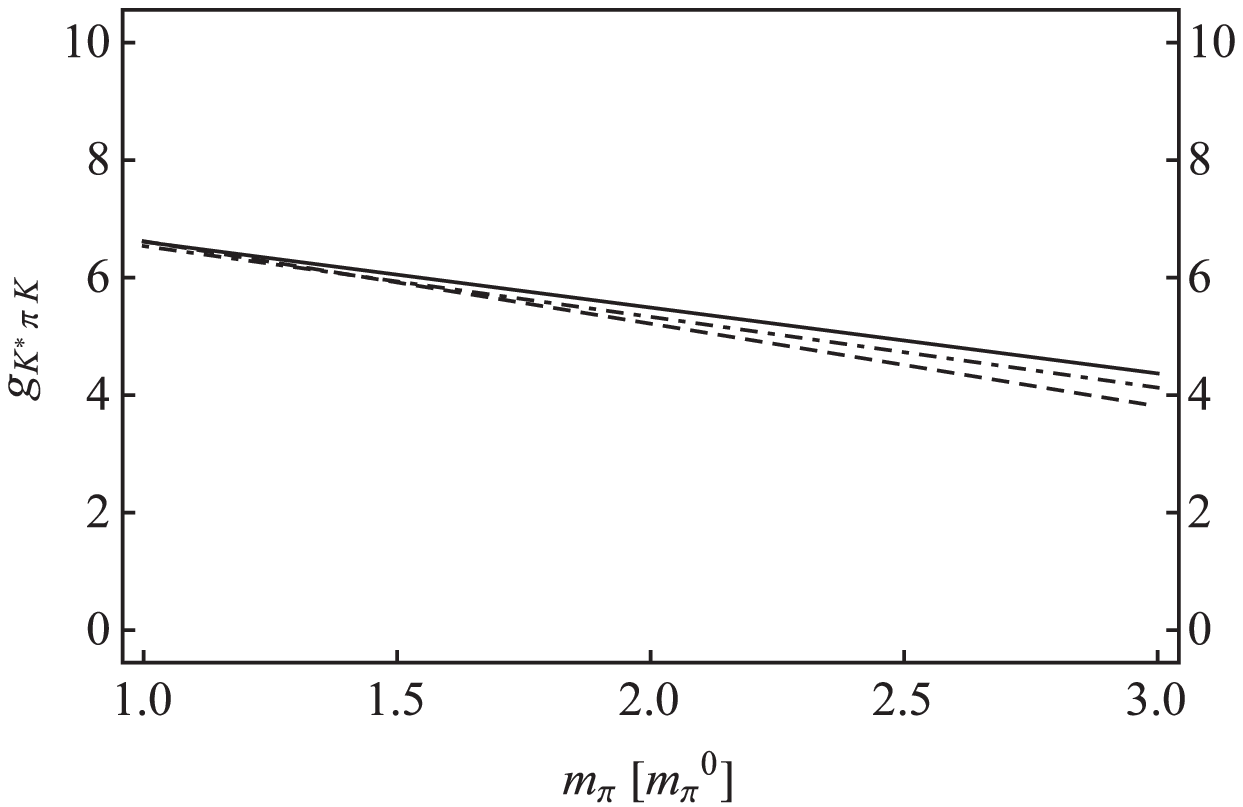}}
\scalebox{0.45}{\includegraphics{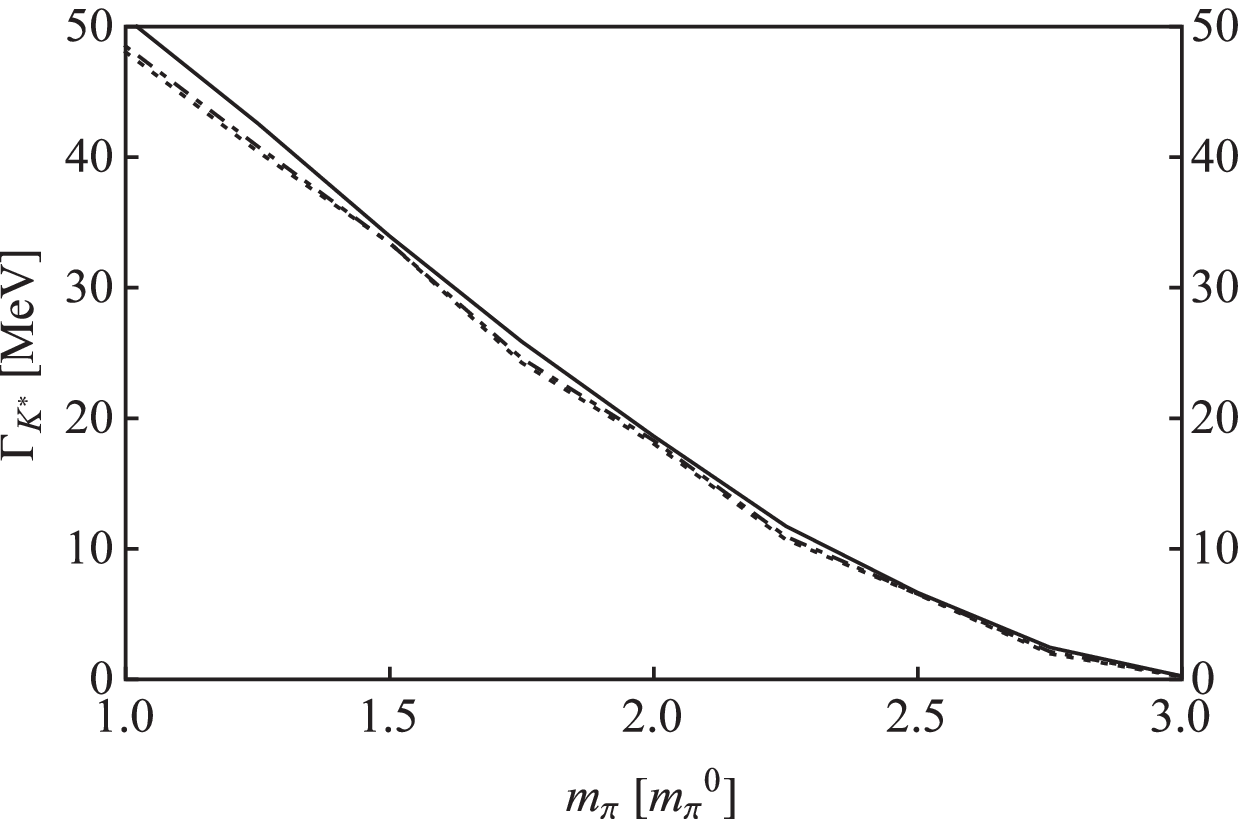}}
\caption{The $K^*$ mass (left), the coupling constant $g_{K^* \pi K}$ (middle) and the decay width $\Gamma_{K^*}$ (right) as functions of $m_\pi$. The solid curves are obtained using the lowest $K \pi$ energy level, the dotdashed curves are obtained using both the first and the second energy levels, and the dashed curves are obtained using the three lowest energy levels. We find that these results are almost the same.} \label{fig:Kstarphy}
\end{center}
\end{figure}

To be complete, we also show the results calculated by fitting the phase shifts obtained using the three lowest energy levels of Fig.~\ref{fig:Kstarphy}. The solid curves are obtained using the first (lowest) $K \pi$ energy level, the dotdashed curves are obtained using both the first and the second energy levels, and the dashed curves are obtained using all the three lowest energy levels. We find that these results are almost the same.

\section{Comparison of our results with the standard L\"uscher's approach}
\label{sec:luscher}

To make our analysis complete, we compare our results with the Lattice results obtained using the standard L\"uscher's approach~\cite{Doring:2011vk}. To do this, we follow the same approach of Refs.~\cite{Chen:2012rp,Boucaud:2007uk,Beane:2007xs,Noaki:2008gx,Pelaez:2010fj}: we use as input the energy levels which we have calculated in Sec.~\ref{sec:energylevels} and Sec.~\ref{sec:pimass} using our $G$-functions (Eqs.~(\ref{eq:GDR}) and (\ref{gtilde})), but evaluate phase shifts using both our method and the L\"uscher's $G$-function.

\begin{figure}[hbt]
\begin{center}
\scalebox{0.65}{\includegraphics{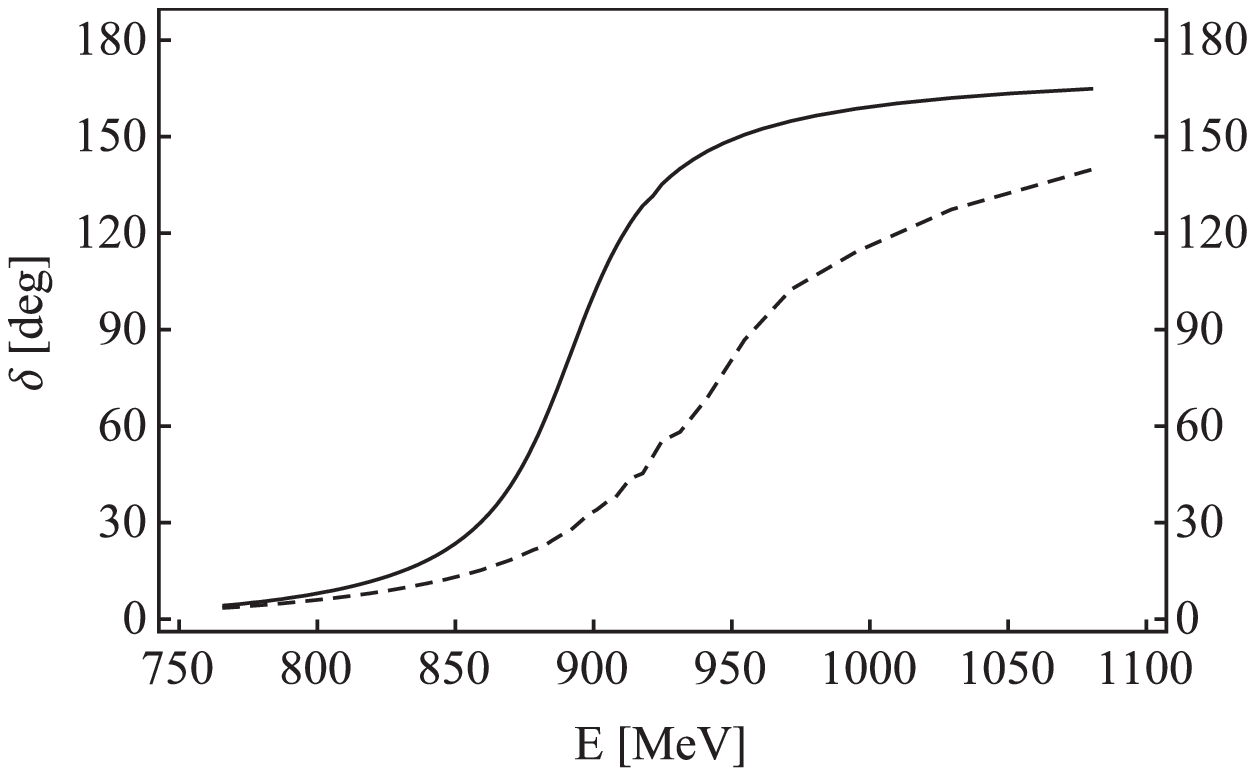}}
\scalebox{0.65}{\includegraphics{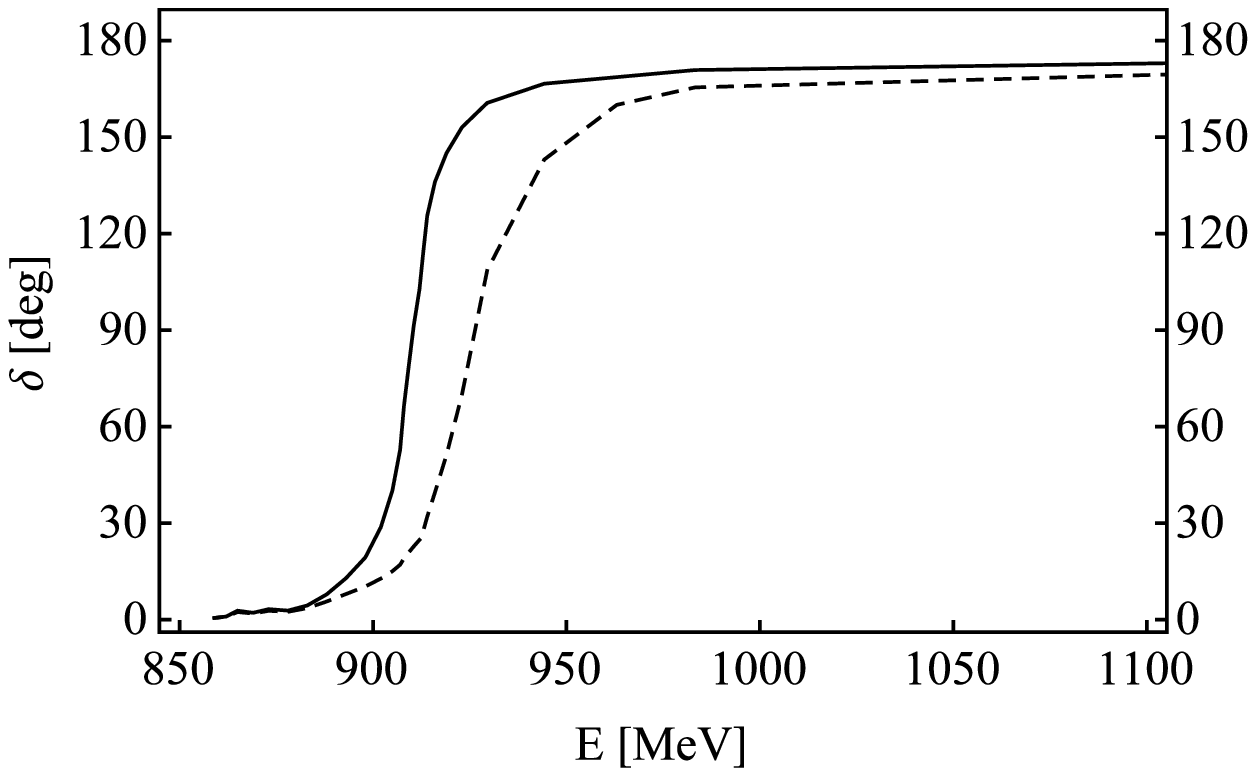}}
\caption{The comparison of our results with results obtained from the standard L$\ddot{\rm u}$scher's approach. The dashed curves are the phase shift evaluated using the standard L$\ddot{\rm u}$scher's approach, while the solid curves are our results. The left panel is obtained using the lowest $K \pi$ energy level shown in Fig.~\ref{fig:Level}, where $m_\pi = 138$ MeV and $m_K = 496$ MeV, while the right panel is obtained using the lowest $K \pi$ energy level shown in the right panel of Fig.~\ref{fig:Lattice}, where $m_\pi = 266$ MeV and $m_K = 552$ MeV~\cite{Lang:2012sv}.}
 \label{fig:compare Luscher}
\end{center}
\end{figure}

The function $I(q_i)$ of Eq.~(\ref{ifun}) can be written as
\begin{equation}
\begin{split}
\label{luscherap}
{1 \over 2 \omega_1 \omega_2} {\omega_1 + \omega_2 \over E^2 - (\omega_1 + \omega_2)^2 + i \epsilon} =~&{1 \over 2E} { 1 \over p^2 - \vec q^2 + i \epsilon }~~~~~~~~~~~~~(a)\\
&- {1 \over 2 \omega_1 \omega_2} {1 \over \omega_1 + \omega_2 + E}~~~~~(b) \\
&- {1 \over 4 \omega_1 \omega_2} {1 \over \omega_1 - \omega_2 - E}~~~~~(c) \\
&- {1 \over 4 \omega_1 \omega_2} {1 \over \omega_2 - \omega_1 - E}~~~~~(d) \, ,
\end{split}
\end{equation}

In the standard L\"uscher's approach only the first term of this equation is kept. We use the following two sets of energy levels: a) the lowest $K \pi$ energy level shown in Fig.~\ref{fig:Level}, where $m_\pi = 138$ MeV and $m_K = 496$ MeV are used; and b) the lowest $K \pi$ energy level shown in the right panel of Fig.~\ref{fig:Lattice}, where $m_\pi = 266$ MeV and $m_K = 552$ MeV are used~\cite{Lang:2012sv}. The obtained phase shifts are shown in Fig.~\ref{fig:compare Luscher}, where the dashed curves are the phase shift evaluated using the standard L$\ddot{\rm u}$scher's approach and the solid curves are our results. These phase shifts can be similarly used to fit the physical quantities:
\begin{eqnarray}
&a):&~~~~m_{K^*} = 961.54 {\rm~MeV}\, , g_{K^* \pi K}=8.25 \, , \Gamma_{K^*} = 110.77 {\rm~MeV} \, ,\\
\nonumber &b):&~~~~m_{K^*} = 926.04 {\rm~MeV}\, , g_{K^* \pi K}=6.48 \, , \Gamma_{K^*} = 17.14 {\rm~MeV} \, ,
\label{lushermass1}
\end{eqnarray}
Comparing these two figures, we clearly see that the L$\ddot{\rm u}$scher's results and our results are quite similar in case (b) when nonphysical pion and kaon masses are used. This confirms the validity of the standard
L\"uscher's approach in the real simulation. There are some differences in case (a) when the physical pion and kaon masses are used, but still both results are consistent with each other within uncertainties, considering the uncertainties related to phase shifts are quite large, as shown in the right panel of Fig.~\ref{fig:certainty}. 

\begin{figure}[hbt]
\begin{center}
\scalebox{0.65}{\includegraphics{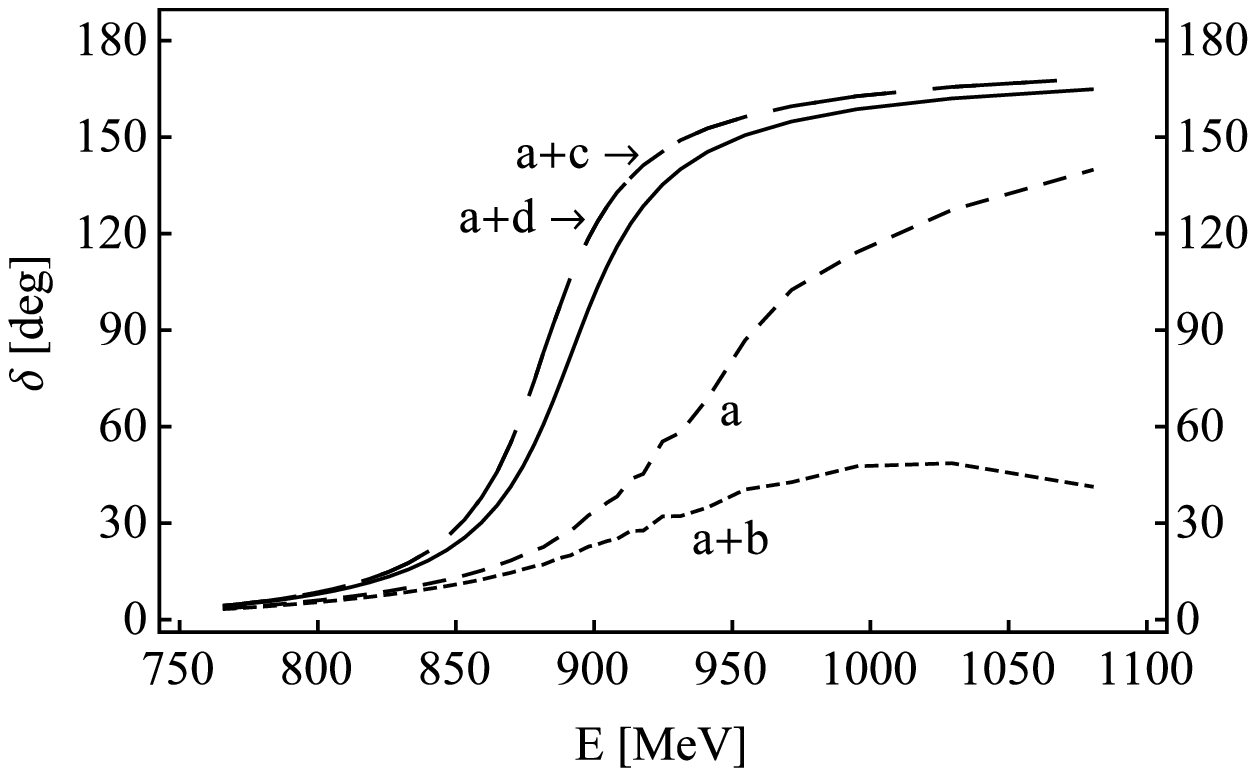}}
\scalebox{0.65}{\includegraphics{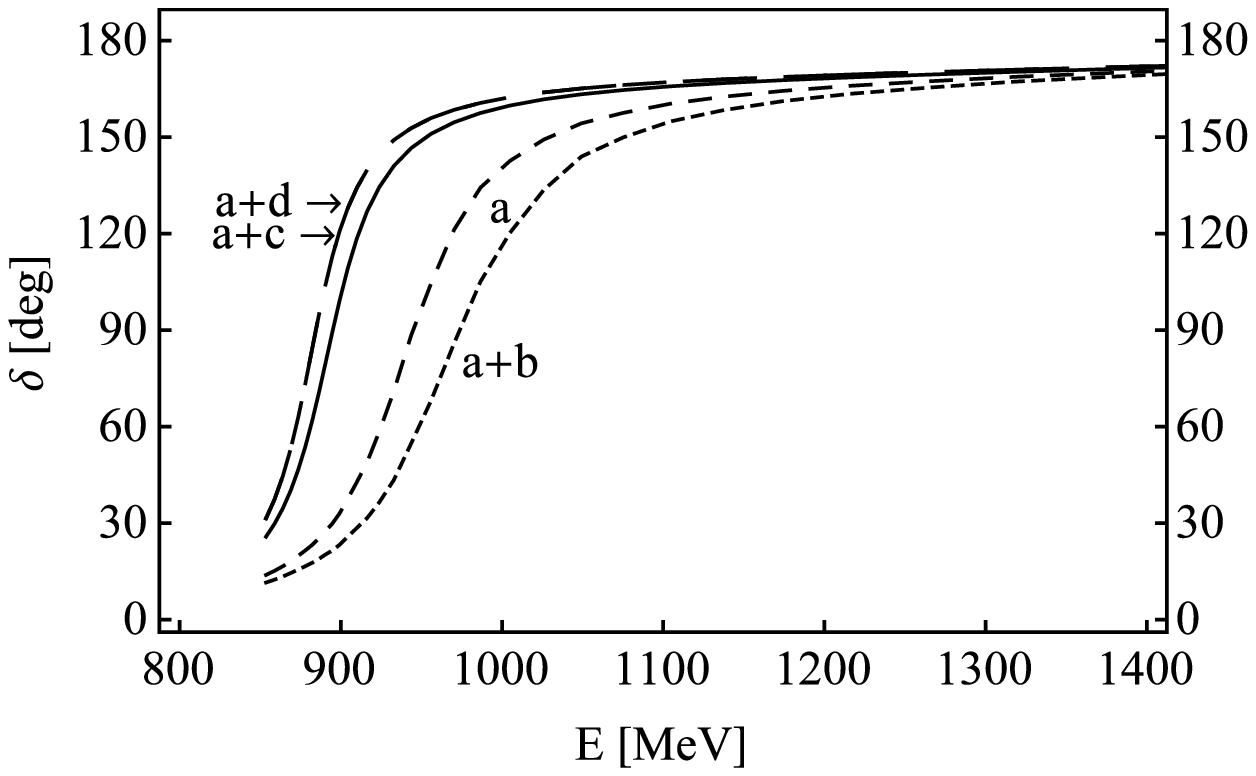}}
\caption{Contributions of the second, the third and the fourth terms of Eq.~(\ref{luscherap}), where physical pion and kaon mass are used, $m_\pi = 138$ MeV and $m_K = 496$ MeV. The solid curves are our results; the dashed curves are the phase shift evaluated using the standard L$\ddot{\rm u}$scher's approach, i.e., the first term of Eq.~(\ref{luscherap}); the short-dashed curves are obtained using the first and the second terms of Eq.~(\ref{luscherap}); the middle-dashed curves are obtained using the first and the third terms of Eq.~(\ref{luscherap}); the long-dashed curves are obtained using the first and the fourth terms of Eq.~(\ref{luscherap}). We note that sometimes the middle-dashed and long-dashed curves are overlapped with each other. In the left panel we show phase shifts extracted from the first energy level of Fig.~\ref{fig:Level}, and in the right panel we show phase shifts obtained using the second energy level of Fig.~\ref{fig:Level}.}
\label{fig:LuscherP}
\end{center}
\end{figure}

\begin{figure}[hbt]
\begin{center}
\scalebox{0.65}{\includegraphics{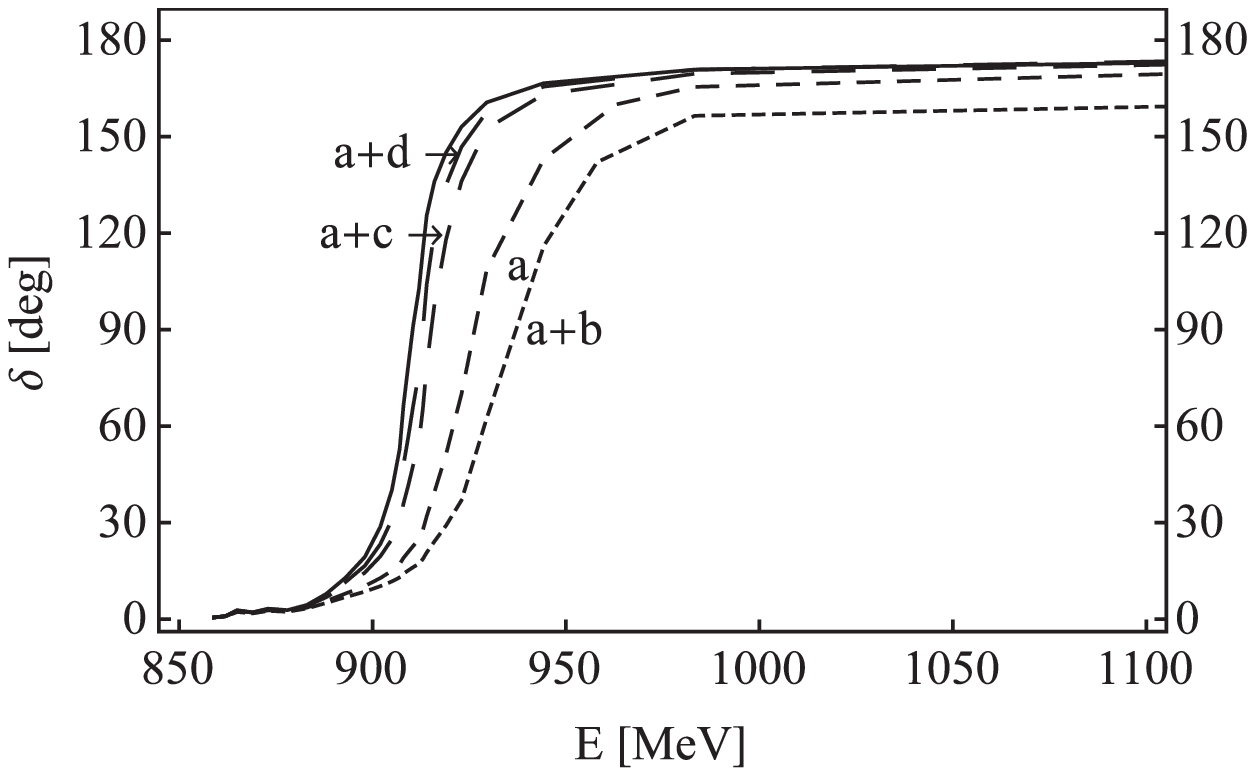}}
\scalebox{0.65}{\includegraphics{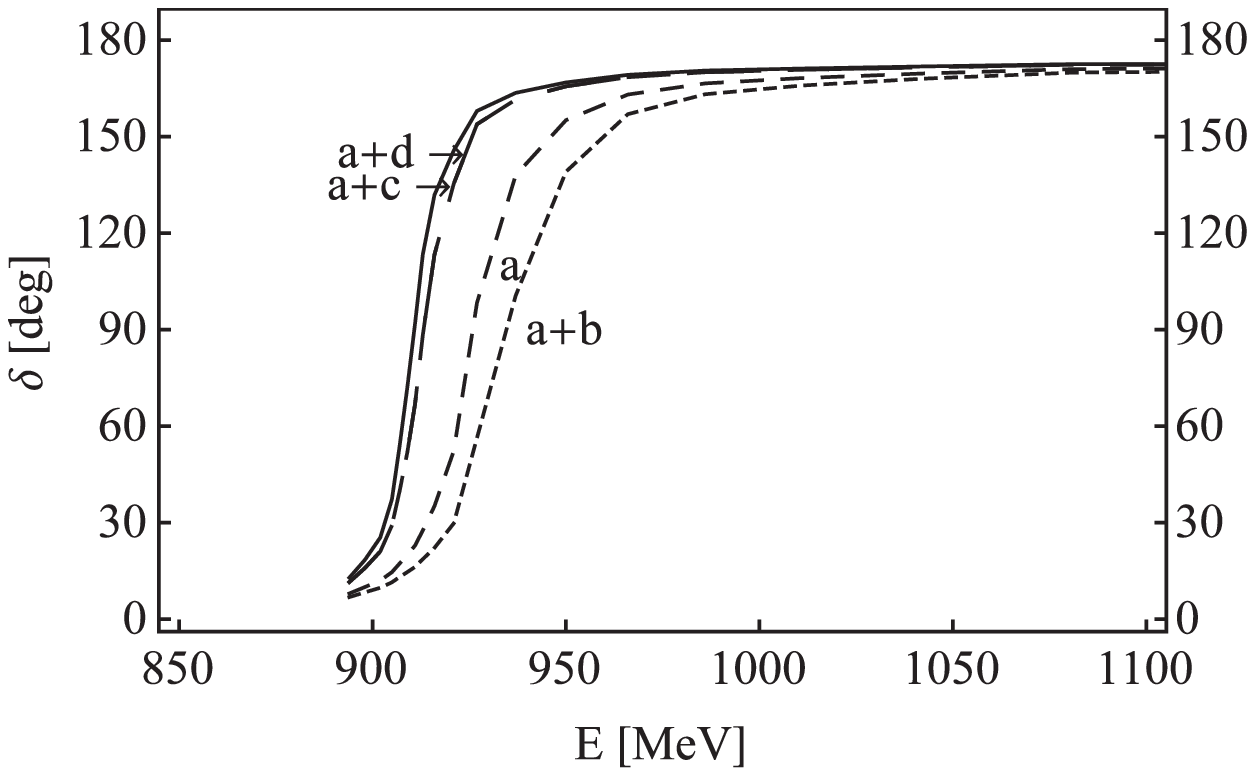}}
\caption{Contributions of the second, the third and the fourth terms of Eq.~(\ref{luscherap}), where non-physical pion and kaon mass are used, $m_\pi = 266$ MeV and $m_K = 552$ MeV~\cite{Lang:2012sv}. We note that sometimes the middle-dashed and long-dashed curves are overlapped with each other.}
\label{fig:LuscherNP}
\end{center}
\end{figure}

We note that this discrepancy is partly caused by hidden systematics in different approaches.
Moreover, the result at higher energy is obtained using a smaller volume, which could cause sizable finite volume effects (see also discussions in Refs.~\cite{Luscher:1990ux,Chen:2012rp}). Accordingly, we would like to suggest Lattice theorists to pay attention to this effect when they extract the physical information using the Lattice data calculated in a (too) small box, for example, if they want to use the physical pion mass but still the computational power is limited.
For completeness, we try to find where these differences come from by adding the second, the third and the fourth terms of Eq.~(\ref{luscherap}) to the first standard L\"uscher's term. The results for physical pion and kaon masses, are shown in Fig.~\ref{fig:LuscherP},
and the results for $m_\pi=266$ MeV and $m_K=552$ MeV, are shown in Fig.~\ref{fig:LuscherNP}. These results suggest that the relativistic corrections can be well taken into account by simply adding either the third or the fourth terms of Eq.~(\ref{luscherap}).

\section{THE INVERSE PROBLEM OF GETTING PHASE SHIFT FROM LATTICE DATA}

In this section we study the inverse process of getting phase shifts from Lattice Data using two energy levels and a parametrized potential. This has also been done in Refs.~\cite{Doring:2011vk,MartinezTorres:2011pr,Doring:2011nd,Xie:2012np,Chen:2012rp,Roca:2012rx}, showing this method is rather efficient. To do this we assume that the first and second energy levels shown in Fig.~{\ref{fig:Level}} are ``Lattice'' inputs, or ``synthetic'' data. We shall use them to inversely evaluate the $V$-matrix and then calculate phase shifts. At the same time we shall give these ``lattice data'' some error bars which can be used to evaluate the uncertainties of the phase shifts.

\begin{figure}[hbt]
\begin{center}
\scalebox{0.62}{\includegraphics{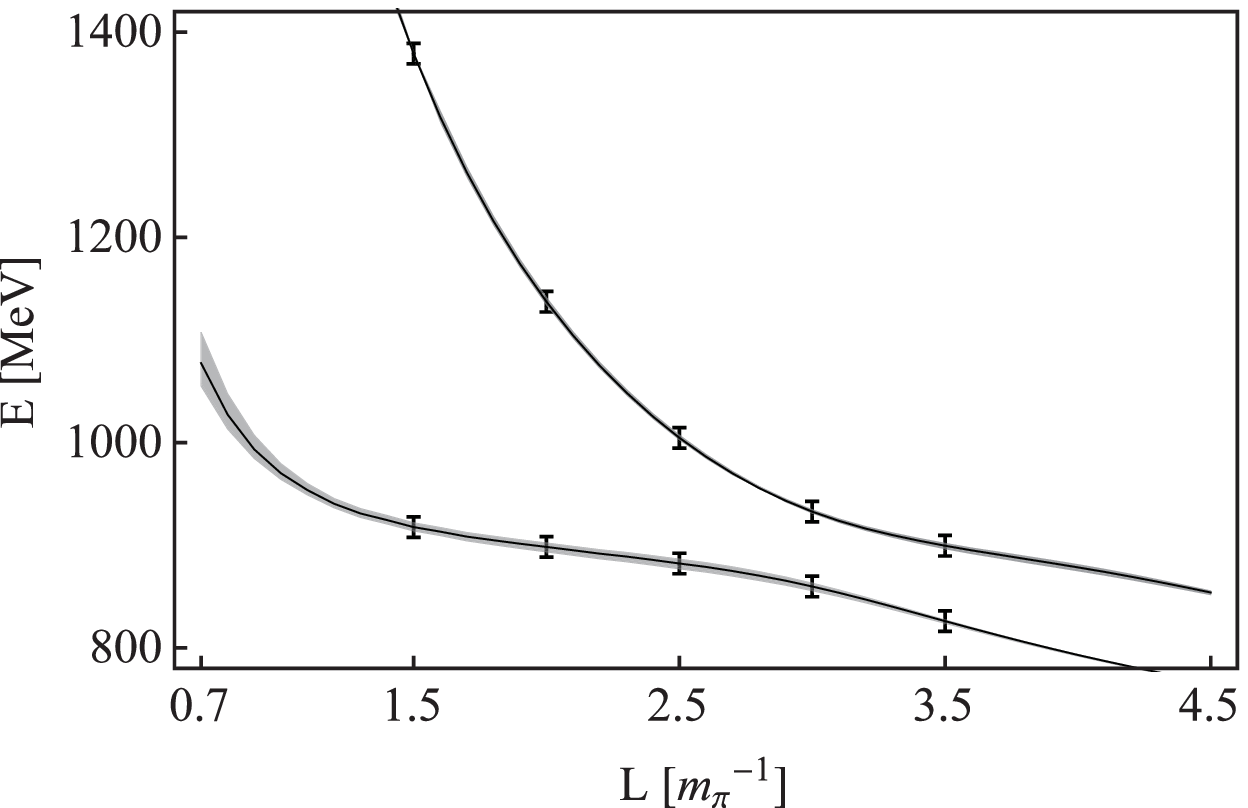}}
\scalebox{0.65}{\includegraphics{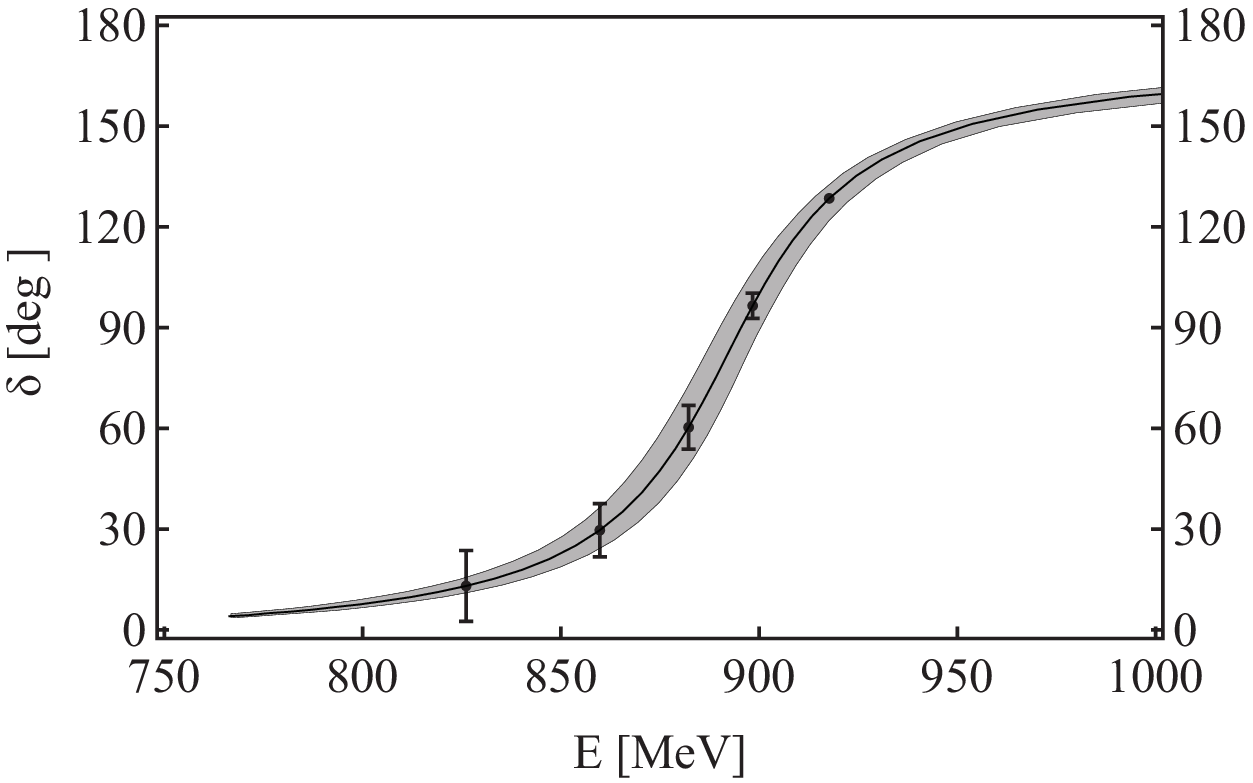}} \caption{The solid curves are the best fitted results $\chi^2_{\rm min} = 0.064$  and the bands are fits that fulfill the $\chi^2 < \chi^2_{\rm min} + 1$ criterion. We note that these bands are not error bars, so it is ok that the right-most point in the right panel is within them. The discrete points on the right figure are the results of the direct determination from each ``data'' points on the left figure.}  \label{fig:fitting}
\end{center}
\end{figure}

Our procedures follow Refs.~\cite{Doring:2011vk,MartinezTorres:2011pr,Doring:2011nd,Xie:2012np,Chen:2012rp,Roca:2012rx}. We take five energies from the first level and five more from the second one (we note that their volumes are also different), and associate to them an error of 10 MeV. Then we use the following function which accounts for a CDD pole \cite{Castillejo:1955ed} to do the one-channel fitting:
\begin{equation}
V=-ap^2(1+\frac{bs}{c^2-s}) \, .
\end{equation}
where $a$, $b$ and $c$ are three free parameters which we shall fit with the ``Lattice'' data shown in Fig.~\ref{fig:Level}. The results are shown Fig.~\ref{fig:fitting} where the energy levels are calculated from all the possible sets of parameters having $\chi^2 < \chi^2_{\rm min} + 1$. Here $\chi^2_{\rm min} = 0.064$ is the best fitting we obtained, where the three parameters are:
\begin{eqnarray}
a= 6.50\times10^{-5} ~{\rm MeV}^{-2},~~~b =0.79,~~~c=918.90 \rm ~MeV  \, ,
\end{eqnarray}
we find that errors in the phase shift are large at small energies, but they become smaller as the energy increases.

As mentioned in Ref.~\cite{Doring:2011vk} the result of this inverse analysis does not depend on which cut off, or subtraction constant one uses in the analysis, as far as one uses the same ones to induce $V$ from the lattice data and then later on to get phase shifts in the infinite volume from Eq. (\ref{bethesal}). The method proves to be practical and efficient.

\section{Conclusion}
\label{sec:summary}

In this paper we use the efficient strategy proposed in Ref.~\cite{Doring:2011vk} to obtain $K \pi$ phase shifts, and thus the $K^*$ meson properties from energy levels obtained in lattice calculations. To do this we studied the $K \pi$ interaction in $P$-wave in a finite box using the chiral unitary approach which has been very successful to provide $K \pi$ phase shifts in infinite space. We evaluated energy levels which are functions of the cubic box size $L$ and the pion mass $m_\pi$. Then we use these energy levels to obtain $K \pi$ phase shifts. Finally we use these phase shifts to fit the physical quantities for the $K^*$ meson: $m_{K^*} = 894.89_{-37.77}^{+39.75} {\rm~MeV}$, $g_{K^* \pi K}=6.48_{-0.12}^{+0.13}$, $\Gamma_{K^*} = 50.68_{-8.00}^{+8.24} {\rm~MeV}$.
To compare our results with the Lattice QCD calculations, we also used non-physical pion masses and redid the same calculations. We note that other parameters can also change with $m_\pi$, and we have considered these effects. The comparison of our results with the Lattice QCD results are shown in Table~\ref{tab:comparedata1} and Table~\ref{tab:comparedata2}, where we can see our results compare favorably with those lattice results obtained in Refs.~\cite{Fu:2012tj,Lang:2012sv}. We note that in these calculations we have estimated the theoretical uncertainties.

To make our analysis complete, we compare our results with the Lattice results obtained using the standard L\"uscher's approach. We find that the L$\ddot{\rm u}$scher's results and our results are quite similar for the $K^*(892)$ resonance when the non-physical pion and kaon masses, $m_\pi = 266$ MeV and $m_K = 552$ MeV, are used. There are some differences when the physical pion and kaon masses are used, although both results are still consistent with each other within uncertainties. This discrepancy is partly caused by hidden systematics in different approaches. Moreover, the result at large energy are obtained using a small volume, and could cause sizable finite volume effects. Accordingly, we would like to suggest Lattice theorists to pay attention to this effect when they extract the physical information using the Lattice data calculated in a (too) small box, for example, if they want to use the physical pion mass but still the computational power is limited. Our analyses also suggest that the relativistic corrections can be well taken into account by simply adding either the third or the fourth term of Eq.~(\ref{luscherap}). We also studied the inverse process of getting phase shifts from our ``synthetic¡¯¡¯ lattice data using two energy levels and a parametrized potential.

\section{Acknowlegement}

We thank Eulogio Oset for suggesting this problem and valuable help, and Bao-Xi Sun, Chu-Wen Xiao and Xiu-Lei Ren for useful discussion and information. We also than Zhi-Hui Guo for useful discussion and information about Ref.~\cite{Guo:2011pa}, and we have started to study the $K^*$(892) resonance in finite volume with coupled channels $K \pi$, $K \eta$ and $K \eta'$. This work is supported by the National Natural Science Foundation of China under Grant No.11205011, 11475015, 11375024 and 11375023, and the Fundamental Research Funds for the Central Universities.


\begin{thebibliography}{99}

\bibitem{luscher}
  M.~L\"uscher,
  Commun.\ Math.\ Phys.\  {\bf 105} (1986) 153.

\bibitem{Luscher:1990ux}
  M.~L\"uscher,
  Nucl.\ Phys.\  B {\bf 354} (1991) 531.

\bibitem{Bernard:2008ax}
  V.~Bernard, M.~Lage, U.~G.~Meissner and A.~Rusetsky,
  JHEP {\bf 0808} (2008) 024.

\bibitem{Bernard:2010fp}
  V.~Bernard, M.~Lage, U.-G.~Meissner and A.~Rusetsky,
  JHEP {\bf 1101} (2011) 019.

\bibitem{Doring:2011vk}
  M.~Doring, U.~-G.~Meissner, E.~Oset and A.~Rusetsky,
  Eur.\ Phys.\ J.\ A {\bf 47}, 139 (2011).

\bibitem{Doring:2011ip}
  M.~Doring, J.~Haidenbauer, U.~-G.~Meissner, A.~Rusetsky,
  Eur. Phys. J. A {\bf 47}, 163 (2011).

\bibitem{MartinezTorres:2011pr}
  A.~Martinez Torres, L.~R.~Dai, C.~Koren, D.~Jido and E.~Oset,
  Phys.\ Rev.\  D {\bf 85}, 014027 (2012).

\bibitem{Kolomeitsev:2003ac}
  E.~E.~Kolomeitsev, M.~F.~M.~Lutz,
  Phys.\ Lett.\  {\bf B582}, 39-48 (2004).

\bibitem{Hofmann:2003je}
  J.~Hofmann, M.~F.~M.~Lutz,
  Nucl.\ Phys.\  {\bf A733}, 142-152 (2004).

\bibitem{Guo:2006fu}
  F.~-K.~Guo, P.~-N.~Shen, H.~-C.~Chiang, R.~-G.~Ping, B.~-S.~Zou,
  Phys.\ Lett.\  {\bf B641}, 278-285 (2006).

\bibitem{Gamermann:2006nm}
  D.~Gamermann, E.~Oset, D.~Strottman, M.~J.~Vicente Vacas,
  Phys.\ Rev.\  {\bf D76}, 074016 (2007).

\bibitem{Doring:2011nd}
  M.~Doring, U.~G.~Meissner,
  JHEP {\bf 1201}, 009 (2012).

\bibitem{Xie:2012np}
  J.~-J.~Xie and E.~Oset,
  Eur.\ Phys.\ J.\ A {\bf 48}, 146 (2012).

\bibitem{Roca:2012rx}
  L.~Roca and E.~Oset,
  Phys.\ Rev.\ D {\bf 85}, 054507 (2012).

\bibitem{Chen:2012rp}
  Hua-Xing ~Chen and E.~Oset,
  Phys.\ Rev.\ D {\bf 87},016014 {2013}

\bibitem{Fu:2012tj}
  Ziwen Fu and Kan Fu,
  Phys.\ Rev.\ D {\bf 86}, 094507 (2012).

\bibitem{Lang:2012sv}
  C.~B.~Lang, and Leskovec Luka and Mohler Daniel and Prelovsek Sasa,
  Phys.\ Rev.\ D {\bf 86}, 054508 (2012).

\bibitem{Prelovsek:2013ela}
  S.~Prelovsek, L.~Leskovec, C.~B.~Lang and D.~Mohler,
  Phys.\ Rev.\ D {\bf 88}, no. 5, 054508 (2013).

\bibitem{Dudek:2014qha}
  J.~J.~Dudek {\it et al.}  [Hadron Spectrum Collaboration],
  Phys.\ Rev.\ Lett.\  {\bf 113}, no. 18, 182001 (2014).

\bibitem{Oller:1998zr}
  J.~A.~Oller and E.~Oset,
  Phys.\ Rev.\ D {\bf 60}, 074023 (1999).

\bibitem{Xiao:2013mn}
  C.~W.~Xiao, F.~Aceti, and M.~Bayar,
  Eur.\ Phys.\ J.\  A {\bf 49}, 22 (2011).

\bibitem{Oller:1997ti}
  J.~A.~Oller and E.~Oset,
  Nucl.\ Phys.\ A {\bf 620}, 438 (1997)
  [Erratum-ibid.\ A {\bf 652}, 407 (1999)].

\bibitem{Oller:2001fj}
  J.~A.~Oller and U.~G.~Meissner,
  Phys.\ Lett.\ B {\bf 500}, 263 (2001).

\bibitem{Altenbuchinger:2013gaa}
  M.~Altenbuchinger and L.~S.~Geng,
  Phys.\ Rev.\ D {\bf 89}, no. 5, 054008 (2014).

\bibitem{Guo:2011pa}
  Z.~H.~Guo and J.~A.~Oller,
  Phys.\ Rev.\ D {\bf 84}, 034005 (2011).

\bibitem{sasa}
  C.~B.~Lang, D.~Mohler, S.~Prelovsek and M.~Vidmar,
  Phys.\ Rev.\ D {\bf 84}, 054503 (2011).

\bibitem{Oller:2000ug}
  J.~A.~Oller, E.~Oset and J.~E.~Palomar,
  Phys.\ Rev.\ D {\bf 63}, 114009 (2001).

\bibitem{Doring:2011jh}
  M.~Doring, J.~Haidenbauer, U.-G.~Meisssner, A.~Rusetsky,
  Eur.\ Phys.\ J.\  A {\bf 47}, 163 (2011).

\bibitem{Estabrooks:1977xe}
  P.~Estabrooks, R.~K.~Carnegie, A.~D.~Martin, W.~M.~Dunwoodie, T.~A.~Lasinski and D.~W.~G.~S.~Leith,
  Nucl.\ Phys.\ B {\bf 133}, 490 (1978).

\bibitem{Mercer:1971kn}
  R.~Mercer, P.~Antich, A.~Callahan, C.~Y.~Chien, B.~Cox, R.~Carson, D.~Denegri and L.~Ettlinger {\it et al.},
  Nucl.\ Phys.\ B {\bf 32}, 381 (1971).

\bibitem{Boucaud:2007uk}
  P.~Boucaud {\it et al.}  [ETM Collaboration],
  Phys.\ Lett.\ B {\bf 650}, 304 (2007).

\bibitem{Beane:2007xs}
  S.~R.~Beane, T.~C.~Luu, K.~Orginos, A.~Parreno, M.~J.~Savage, A.~Torok and A.~Walker-Loud,
  Phys.\ Rev.\ D {\bf 77}, 014505 (2008).

\bibitem{Noaki:2008gx}
  J.~Noaki, S.~Aoki, T.~W.~Chiu, H.~Fukaya, S.~Hashimoto, T.~H.~Hsieh, T.~Kaneko and H.~Matsufuru {\it et al.},
  PoS LATTICE {\bf 2008}, 107 (2008).

\bibitem{Pelaez:2010fj}
  J.~R.~Pelaez and G.~Rios,
  Phys.\ Rev.\ D {\bf 82}, 114002 (2010).

\bibitem{sakurai} J.~J.~Sakurai, Currents and mesons (University of Chicago Press, Chicago Il 1969).

\bibitem{Bando:1987br}
  M.~Bando, T.~Kugo and K.~Yamawaki,
  Phys.\ Rept.\  {\bf 164}, 217 (1988).

\bibitem{Ecker:1989yg}
  G.~Ecker, J.~Gasser, H.~Leutwyler, A.~Pich and E.~de Rafael,
  Phys.\ Lett.\  B {\bf 223}, 425 (1989).

\bibitem{Zhou:2014ila}
  Y.~Zhou, X.~L.~Ren, H.~X.~Chen and L.~S.~Geng,
  Phys.\ Rev.\ D {\bf 90}, no. 1, 014020 (2014).

\bibitem{Ren:2012aj}
  X.~-L.~Ren, L.~S.~Geng, J.~Martin Camalich, J.~Meng and H.~Toki,
  JHEP {\bf 1212}, 073 (2012).

\bibitem{Castillejo:1955ed}
  L.~Castillejo, R.~H.~Dalitz and F.~J.~Dyson,
  Phys.\ Rev.\  {\bf 101}, 453 (1956).


\end{thebibliography}
\end{document}